\documentclass[journal]{IEEEtran}
\usepackage{subfigure}
\usepackage{graphicx}
\usepackage{amsmath}
\usepackage{multicol}
\usepackage{xcolor}
\usepackage{cite}
\begin{document}

\title{A High Count-Rate and Depth-of-Interaction Resolving Single Layered One-Side Readout Pixelated Scintillator Crystal Array for PET Applications}

\author{J.~M.~C.~Brown,
        S.~E.~Brunner
        and~D.~R.~Schaart%
\thanks{J. M. C. Brown is with the Department of Radiation Science and Technology, Delft University of Technology, The Netherlands, and Centre for Medical Radiation Physics, University of Wollongong, Australia. e-mail: jeremy.brown@cern.ch}
\thanks{S.~E.~Brunner and D.~R.~Schaart is with the Department of Radiation Science and Technology, Delft University of Technology, The Netherlands.}

\thanks{}}

\markboth{}%
{}
\maketitle

\begin{abstract}

Organ-specific, targeted Field-of-View (FoV) Positron Emission Tomography (PET)/Magnetic Resonance Imaging (MRI) inserts are viable solutions for a number of imaging tasks where whole-body PET/MRI systems lack the necessary sensitivity and resolution. To meet the required PET detector performance of these systems, high count-rates and effective spatial resolutions on the order of a few mm, a novel two-axis patterned reflector foil pixelated scintillator crystal array design is developed and its proof-of-concept illustrated in-silico with the Monte Carlo radiation transport modelling toolkit Geant4. It is shown that the crystal surface roughness and phased open reflector cross-section patterns could be optimised to maximise either the PET radiation detector's effective spatial resolution, or count rate before event pile up. In addition, it was illustrated that these two parameters had minimal impact on the energy and time resolution of the proposed PET radiation detector design. Finally, it is shown that a PET radiation detector with balance performance could be constructed using ground crystals and phased open reflector cross-section pattern corresponding to the middle of the tested range.

\end{abstract}


\section{Introduction}

\IEEEPARstart{O}{rgan}-specific, targeted FoV PET/MRI inserts are viable solutions for a number of imaging tasks where whole-body PET/MRI systems lack the necessary sensitivity and resolution \cite{Gonzalez2017,Gonzalez2018a,Gonzalez2018b,Hypmed2019}. Whilst these systems' smaller PET bore diameters of approximately 10 to 30 cm result in increased sensitivity across their FoV, it also increases the impact of ``parallax error'' on system spatial resolution that arise from Depth-of-Interaction (DoI) effects within clinical PET radiation detectors \cite{Gonzalez2018a,Lewellen2008,Ito2011}. To suppress these effects, and reach the target spatial resolutions of 1 mm, compact MR compatible photosensors coupled to crystal arrays with adequate x-y and DoI resolution are required without compromising energy resolution, time resolution and maximum count rate \cite{Schaart2016,Bisogni2018}. At present three primary designs of single sided readout scintillator detectors have been developed to solve this issue: mutli-layered pixelated arrays \cite{Yamashita1990,Nagai1999,Liu2001,Zhang2002,Tsuda2004,Inadama2006}, monolithic \cite{Schaart2009,Schaart2010,Borghi2016}, and one/two-axis light sharing patterned reflector foil crystal arrays \cite{Miyaoka1998,Lewellen2004,Ito2010,Ito2013,Lee2015}.

Multi-layered pixelated scintillator PET detectors were the first of these three PET single sided readout scintillator detector types developed \cite{Yamashita1990,Nagai1999}. They were designed with the specific purpose of enabling DoI measurement via an encoded light sharing pattern determined through specific crystal array layer offsets with respect to one another \cite{Liu2001,Tsuda2004,Inadama2006}. With this approach it became possible to identify which scintillator crystal and layer the gamma-ray interacted. However, this functionality also came at the cost of energy and time resolution due to optical photon scattering between crystal interfaces \cite{Ito2011}.

Monolithic scintillator detectors implement a single reflective foil wrapped crystal and utilise the light sharing distribution of optical photons over the whole surface of the spatially resolving optical photosensor to determine gamma-ray interaction location \cite{Schaart2009}. This simple, yet-robust, design resulted in a radiation detector that possesses high energy, x-y spatial, temporal and DoI resolution \cite{Schaart2010}. However, this type of PET detector is not an ideal candidate for all target FoV PET/MR imaging insert applications due to the possible occurrence of saturation effects from the high PET radiotracer concentration in close proximity \cite{Ito2011}.

One/two-axis light sharing patterned reflector foil crystal arrays were first proposed via Miyaoka et al. \cite{Miyaoka1998}, and recently further improved by Ito et al. \cite{Ito2010}. These systems implemented light spreading along specific axis within a crystal array via reflectors that partially cover the crystal surfaces. Ito et al.'s standard triangular pattern on the top and bottom half of the inter crystal foils, along the x- and y-axis respectively, was shown to enable DoI measurement based on the extent of light shared along each axis \cite{Ito2010}. Further exploration of this novel PET detector technology has illustrated that it was able to obtain excellent energy, x-y spatial and modest DoI resolution when coupled to SiPMs \cite{Ito2013,Lee2015}. However, the long-range light-sharing distributions required to yield this DoI information limits their maximum count rate and, in the case of their application to target FoV PET/MR imaging inserts, there is a high probability that they will suffer from event pile-up effects due to high PET radiotracer concentration in close proximity (e.g. heart and liver uptake in target breast cancer imaging).

This work outlines and presents an in-silico proof-of-concept investigation of a novel two-axis patterned reflector foil pixelated scintillator crystal array design intended for organ-specific, targeted FoV PET/MR inserts. A controllable light-sharing approach was developed through a repeating phased open reflector cross-section pattern along each light-sharing axis. This novel design creates virtual light trapping boundaries, a floating light isolating region of crystals within a larger scintillator crystal array, and enables the determination of DoI whilst minimising the probability of event pile-up. An overview of the light-sharing patterned reflector foil array geometry concept is outlined in Section \ref{sec:M}, followed by a description of the in-silico proof-of-concept investigation for a PET radiation detector intended for the breast cancer imaging PET/MR insert HYPMED \cite{Hypmed2019}. The results of this in-silico investigation, their discussion and an overall conclusion are then presented in Sections \ref{sec:R}, \ref{sec:D} and \ref{sec:C} respectively.

\section{Method}
\label{sec:M}
\subsection{Light-Sharing Patterned Reflector Foil Array Geometry}

\begin{figure}[tbh]    
    \centering
        \includegraphics[width=0.5\textwidth]{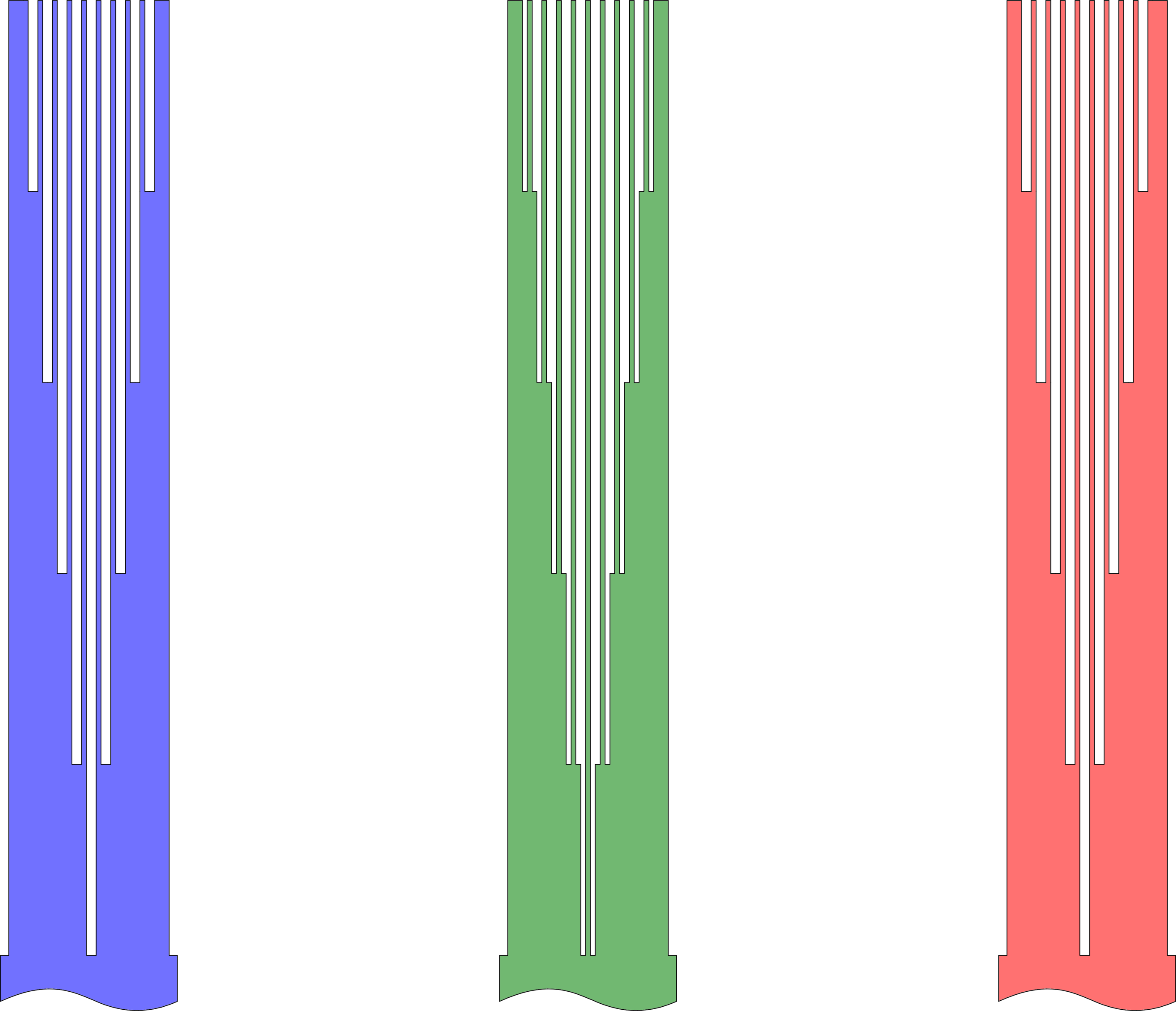}
\caption{Top half of set of 5 step encoded reflective foil designs intended to control the light-sharing to 3 pixelated crystals within a pixelated crystal array. Here, the Phase Shifted Inserts (PSIs) are populated in periodic manner from left to right.}
\label{fig:1}
\end{figure}

Previous single side readout PET radiation detectors have relied on high levels of light sharing within crystals/crystal arrays, and in-turn across the majority of the coupled photosensor surface, to determine the DoI of gamma-rays. This work proposes a novel encoded reflective foil array design in which scintillator crystals are placed to control the extent of light sharing across the array to a desired range, whilst still enabling DoI measurement. These encoded reflective foils possess a step like structure which spans a maximum of half the foil height (z-axis), with the area of each step separated into equally sized sub-regions (see Figure \ref{fig:1}). The number of sub-regions in each step is proportional to the desired light sharing range as a function of the number of pixels (i.e. 3 sub-regions for a desired light sharing range of 3 crystals).

Along each light sharing axis the encoded foils take turns of having one of the the sub-regions filled with a Phase Shifted Insert (PSI) in a periodic manner (i.e. left to right like seen in Figure \ref{fig:1}). These PSI varied reflective foils are placed in repeating pattern perpendicular to the desired direction of light-sharing (i.e. x-axis) and then rotated via 180 degrees before being placed in the same manner along the other direction (y-axis). This linear offset of PSIs in each foil produces a decreasing effective open cross-section with increasing pixelated crystal distance (66\%, 33\%, and 0\% for 1, 2 and 3 crystals spanned respectively), creating a virtual full reflective foil boundary that limits the range of light sharing to a desired distance from the site of gamma-ray interaction ($\pm$ 3 crystals). Furthermore this repeating foil structure enables a unique light sharing distribution along both the x- and y-axis dependent on DoI, and reduces the probability of multiple gamma-rays being detected as a single event due to overlapping light distributions.

\begin{figure}[tbh]   
   \centering
   \includegraphics[width=0.4\textwidth]{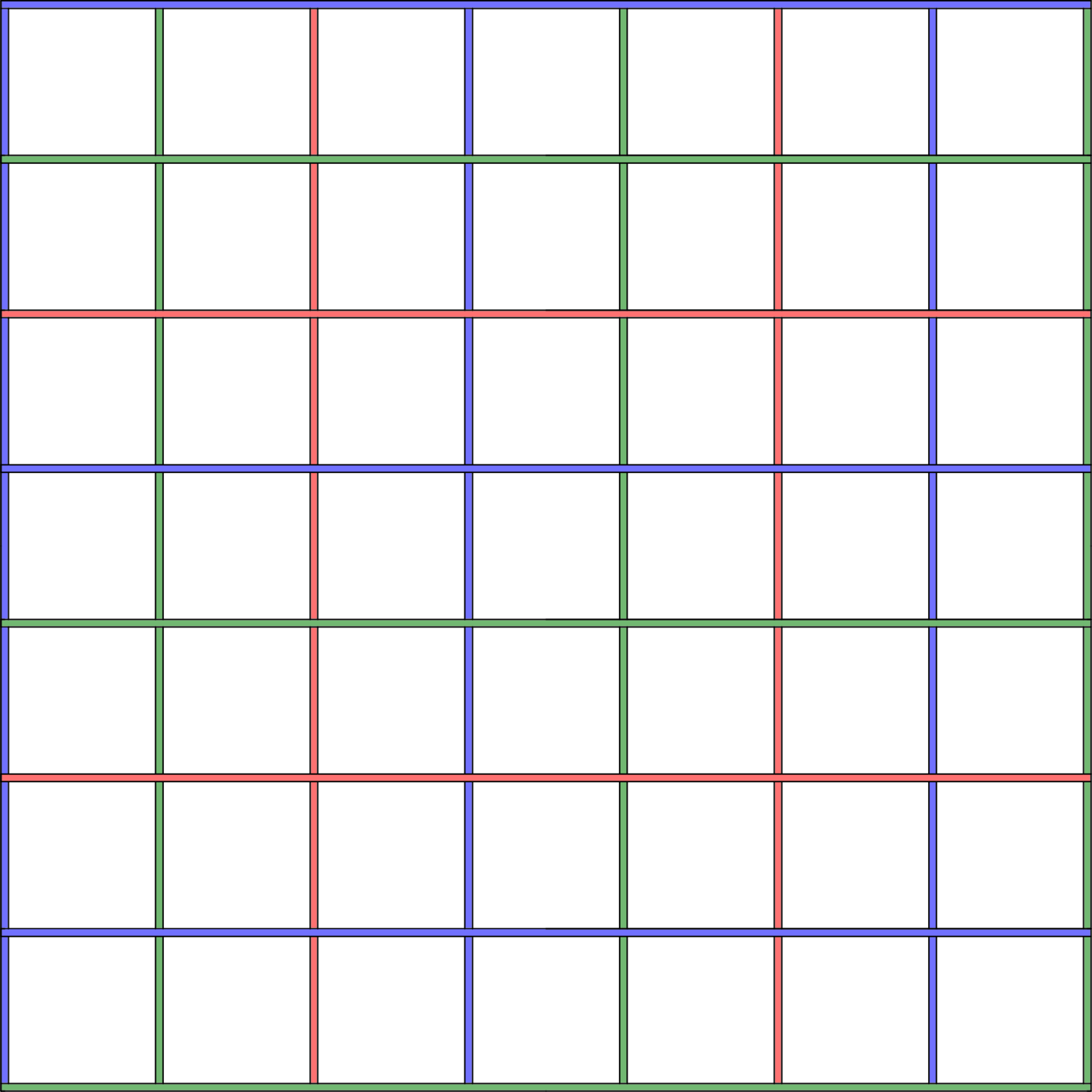}
\caption{A 7 by 7 array of pixelated scintillator crystals constructed to limit light sharing to within a range of 3 pixels from the point of gamma-ray interaction based on the open cross-section designs seen in Fig. \ref{fig:1}. Here the colour coding of each segment corresponds to same for each foil designs seen in Fig \ref{fig:1}, and the x-axis are populated with the foil openings pointing up and the y-axis with the foils pointing down.}
\label{fig:2}
\end{figure}

\begin{figure}[tbh]    
    \centering
    \begin{subfigure}
    \centering 
        \includegraphics[width=0.225\textwidth]{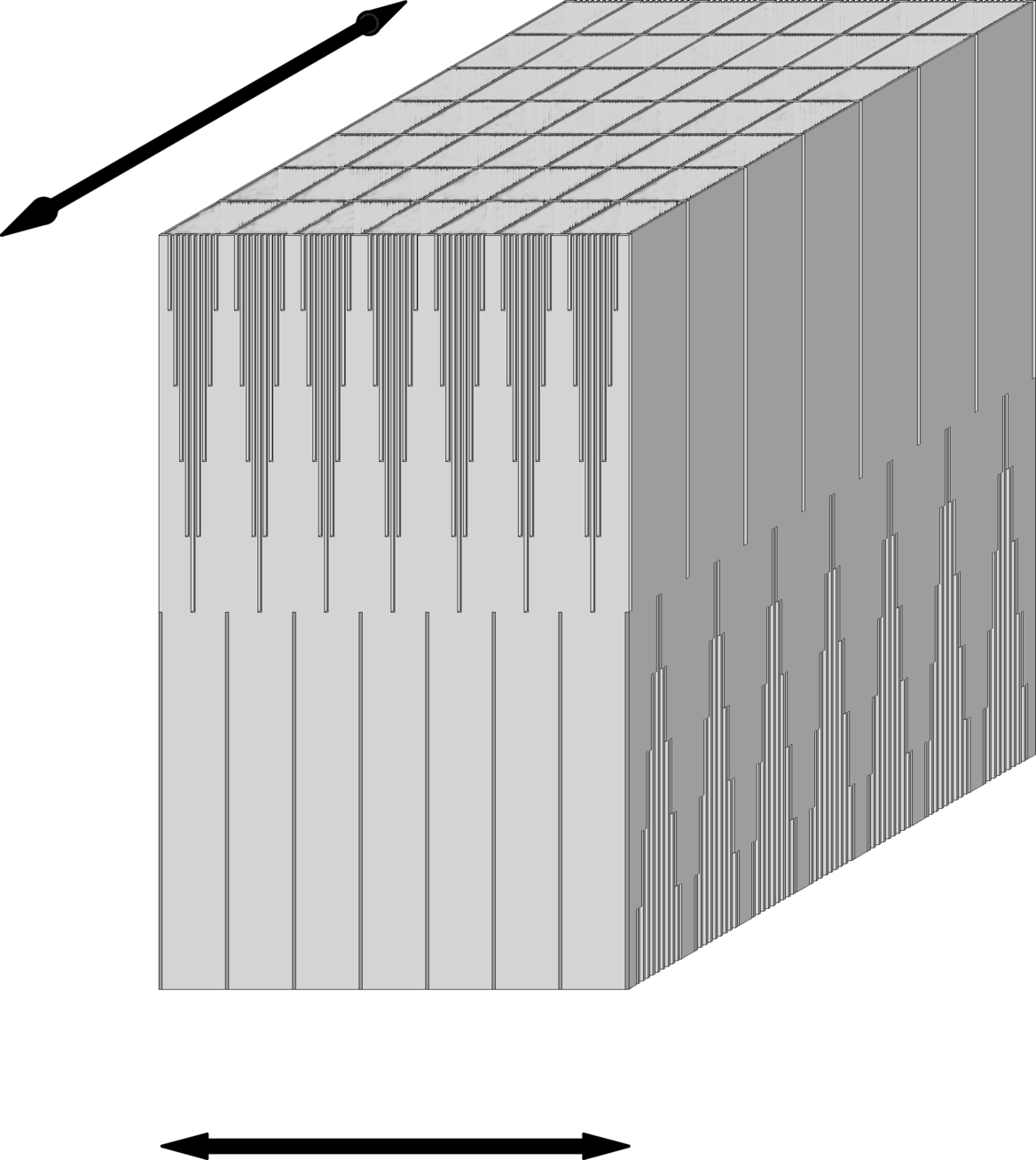}
        \label{fig:3a}
    \end{subfigure}
    \begin{subfigure}
    \centering
        \includegraphics[width=0.225\textwidth]{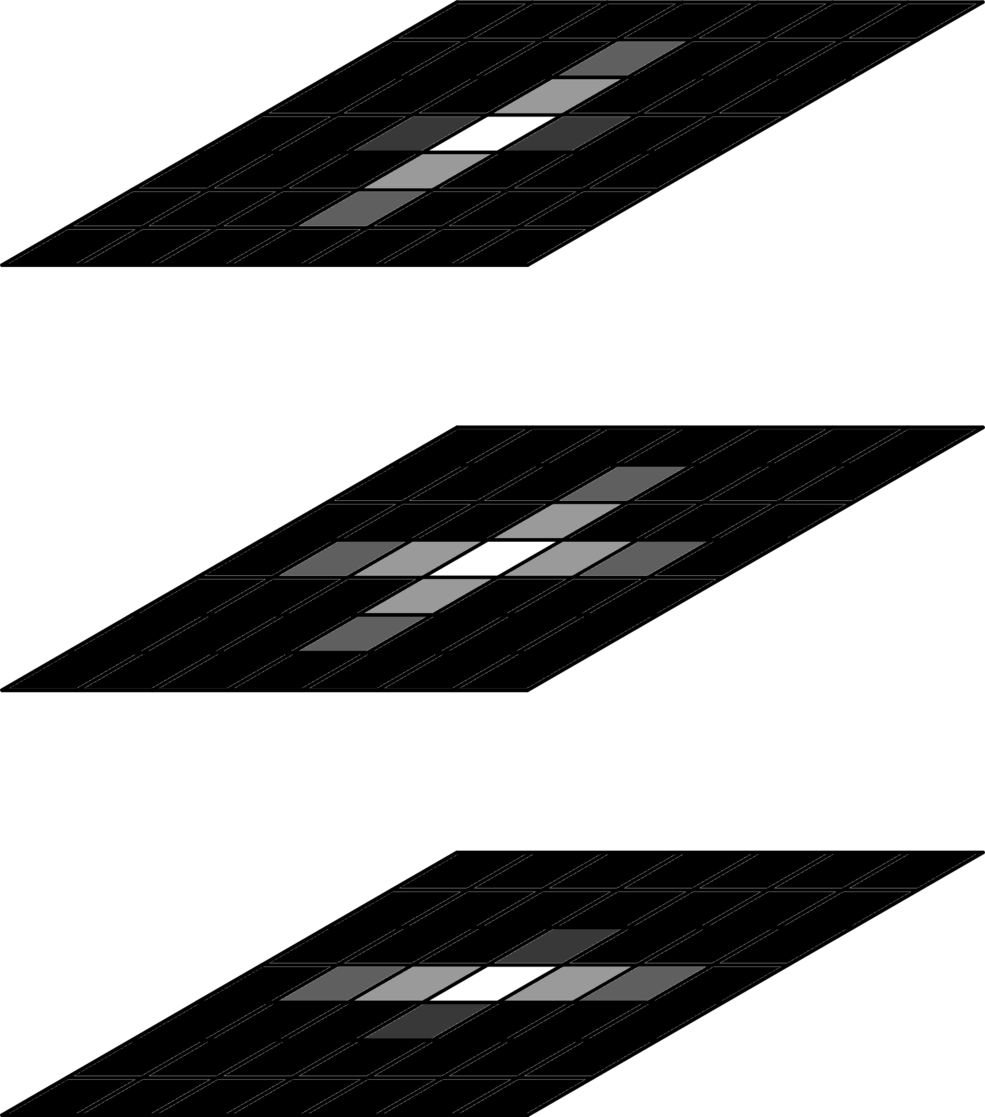}
        \label{fig:3b}
    \end{subfigure}
\caption{Axial light-sharing direction (left) and ideal light-sharing distribution at the bottom surface for the full wrapped array based on the foil designs and population pattern contained in Figures \ref{fig:1} and \ref{fig:2} respectively. Here the light-sharing distributions for gamma-ray photoelectric absorption at the top, middle and bottom within the central crystal can be seen at their respective position (right).}
\label{fig:3}
\end{figure}

An exemplar set of encoded reflective foils designed to limit light sharing to within a range of 3 pixels from the point of gamma-ray interaction is shown in Figure \ref{fig:1} (note: only the top half of the foil is presented). With these foils it is possible to construct a 3 pixelated scintillator crystal range limited light sharing array through the population pattern displayed in Figure \ref{fig:2}, where the x-axis are populated with the foil openings pointing up and the y-axis with the foils pointing down. The colour of each foil segment represents the foil designs seen in Figure \ref{fig:1}. Figure \ref{fig:3}~(left) illustrates the axis of light sharing of this array dependent on interaction height within the pixelated scintillator crystal, out/into the page at the top and across the page at the bottom, and Figure \ref{fig:3}~(right) the ideal light sharing distributions for a full wrapped array bonded to a photosensor.

In Figure \ref{fig:3}~(right) three different interaction depths of gamma-rays within the central pixelated scintillator crystal can be seen: at top, in the middle, and at the bottom close to the optical photosensor. These unique light-sharing distribution along the x- and y-axis illustrate DoI dependence and that their interaction position can be retrieved with an appropriate analysis method ( such as the categorical average pattern nearest-neighbour "Closest Pixel Intensity" \cite{VanDam2011,VanDam2013}, gradient tree-boosted machine learning \cite{Muller2018,Muller2019}, statistically driven maximum-likelihood estimation \cite{Lee2015}, and weighted east squares \cite{Ling2008} approaches), whilst limiting the extent of light-sharing to minimise the probability of multiple gamma-rays being detected as one. Further control over the extent of light-sharing can be obtained by dilating the based width of the PSIs, i.e. equal to the foil step sub-region width, seen in Figure \ref{fig:1}. This PSI dilatation enables for both the maximum range of light-sharing and maximum detector count-rate to be optimised as desired, with full light isolation of each individual crystal occurring when the dilation factor approaches that of the number of sub-regions (e.g. a PSI dilation of 3 for the foil designs outlined in Figure \ref{fig:1}).

\subsection{In-Silico Proof-of-Concept Investigation}

An in-silico test platform was constructed using the Monte Carlo radiation transport modelling toolkit Geant4 version 10.3 \cite{G42003,G42006,G42016} to explore a PET radiation detector design intended for the HYPMED PET/MR insert (http://www.hypmed.eu/). The HYPMED PET/MR insert will be a target breast cancer imaging insert composed of two 160 mm diameter bore PET rings with individual receive-only local RF coils. Previous simulations have shown that for the insert to reach the target PET imagining spatial resolution and sensitivity, of approximately 1 mm and four times that of current clinical systems respectively, a LYSO based radiation detectors with an effective spatial resolution of 1.333 mm, DoI resolution on the order of 5 mm, and crystal thickness of 15 mm is required. The following description of the developed in-silico platform and the proof-of-concept investigation is separated into four primary areas: 1) detector geometry and material, 2) physics and optical surface modelling, 3) photosensor response and PET detector readout modelling, and 4) PET radiation detector performance assessment/optimisation.

\subsubsection{Detector Geometry and Materials}

\begin{figure}[tbh]   
   \centering
   \includegraphics[width=0.4\textwidth]{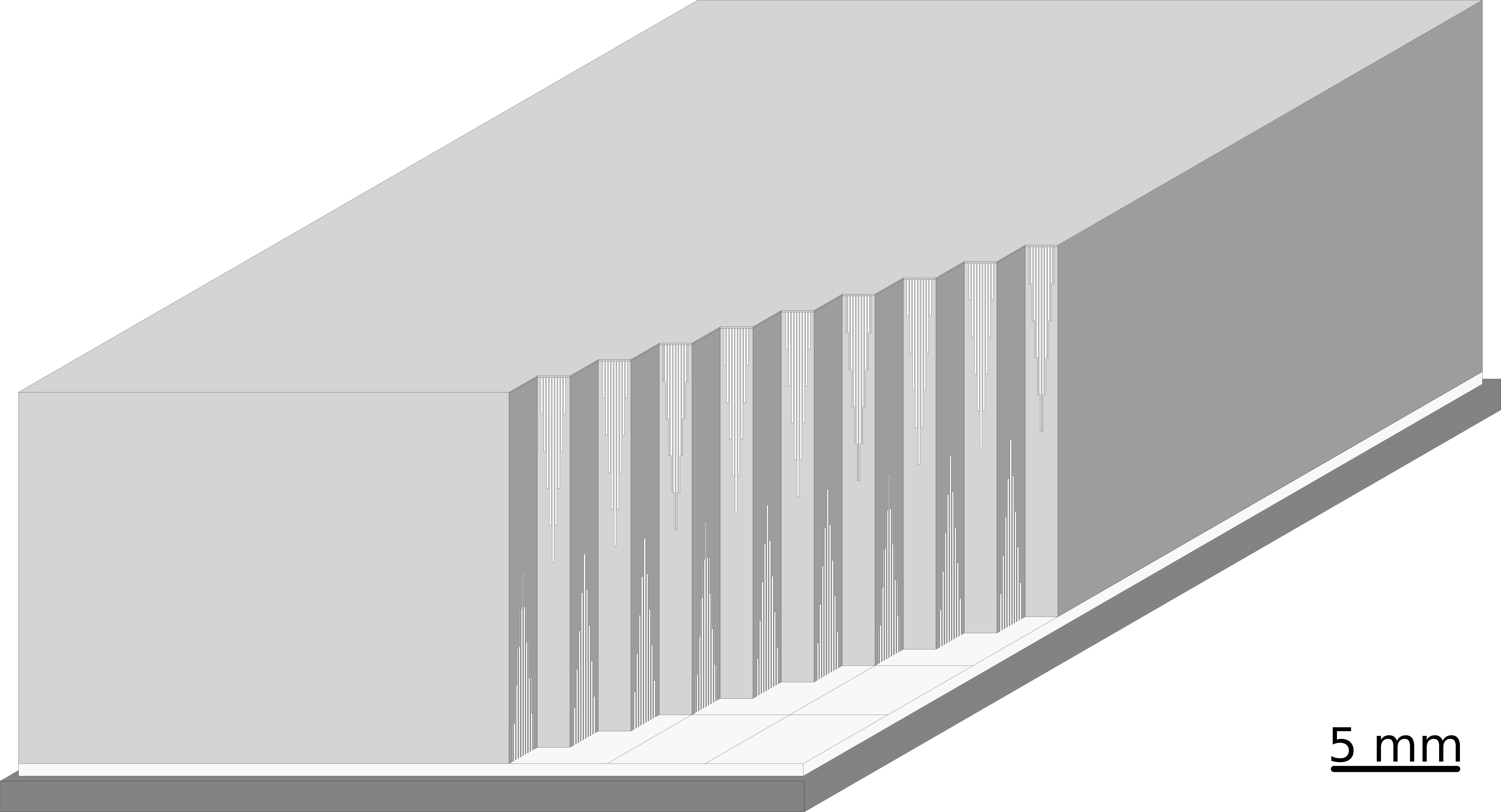}
\caption{A schematic of the PET radiation detector geometry constructed within the Geant4 in-silico test platform. Here a number of crystals have been removed from the 24 by 24 and Vikuiti ESR foils/wapping clipped to illustrate the effective 3$\times$3:1 coupling of LYSO crystals to each SiPM pixel.}
\label{fig:4}
\end{figure}

A schematic of the PET radiation detector composed of a single layered one-side readout pixelated scintillator crystal array, with outer and top wrapping, coupled to a Philips DPC3200 Silicon Photomultiplier (SiPM) \cite{Frach2009,Frach2010} is shown in Figure \ref{fig:4}. The crystal array is composed of an encoded Vikuiti ESR foil separated and wrapped array of 24 by 24 LYSO crystals (1.26 (X)$\times$1.26 (Y)$\times$15.0 (Z) mm) mounted onto the quartz glass protector of a Philips SiPM with a 0.1 mm thick layer of DELO photobond 4436 glue. An identical encoded ESR foil array pattern to that seen in Figures \ref{fig:1}, \ref{fig:2} and \ref{fig:3} was implemented (i.e. a 5 step height, 3 layer repeating PSI structure). These foil parameters were selected to limit the light sharing range to $\pm$ 3 crystals, and in-turn restrict the range of light sharing to a 3$\times$3 SiPM pixel footprint per gamma-ray to increase the potential maximum count-rate before pile-up occurs. 

Within the in-silico test platform regions of open cross-section in the ESR foils between the LYSO crystals were modelled as being filled with air, whereas the outer and top layers were implemented flush to the LYSO crystal surfaces mimicking the process of pressure wrapping to increase structural stability. In the case of the implemented Philips SiPM geometry the layered structure, dimensions and locations of the quartz light guide, glue layers, 8 by 8 array of SiPM pixels, and printed circuit board was taken from version 1.02 of the unit manual \cite{DPCManual2016}. Finally, the density, elemental composition, and optical/scintillation properties of all materials can be found in Appendix \ref{appendix1}.

\subsubsection{Physics and Optical Surface Modelling}

Gamma-ray, x-ray and electron transport was simulated using the Geant4 Option4 EM physics list (G4EmStandardPhysics\_option4 \cite{G42016}) with atomic deexcitation enabled, a maximum particle step length of 10 $\mu$m, and a low-energy cut off of 250 eV. Optical photon generation and transport was included for the processes of scintillation, absorption, refraction and reflection through implementation of the available Geant4 "Unifed" model \cite{Levin1996}. With the exception of the ESR foil to other material interfaces (modelled as a metal to dielectric with reflectivity matching the 3M Vikuiti ESR data sheet), the optical interface of all other materials was modelled as a dielectric to dielectric. Furthermore, all but one material optical interfaces were described as ground surfaces with a surface roughness of 0.1 degrees (i.e. its not possible for surfaces to be ``perfectly smooth") \cite{VanderLaan2010,Nilsson2015}. The singular exception was the surface roughness of four sides of each LYSO crystal which was optimised to maximise the PET radiation detector performance (see Section \ref{sec:m1o}).

\subsubsection{Photosensor Response and PET Detector Readout Modelling}

Modelling of the photosensor response was implemented in two steps: 1) physical geometry, and 2) electronic response. The physical geometry of the Philips DPC3200 SiPM was achieved through the definition of scoring boundaries that mimicked the shape and location of all 3200 59.4 $\mu$m$\times$64 $\mu$m Single Photon Avalanche Diodes (SPADs) \cite{Frach2009,Frach2010} within each SiPM pixel of the Geant4 test platform. As for the electronic behaviour of the photosensor, it was modelled based on four assumptions: 1) the probability of a photoelectrically absorbed optical photon triggering a SPAD is proportional to the Photon Detection Efficiency (PDE) outlined in \cite{Frach2009}, 2) a given SPAD can only trigger once per simulated primary particle (be it gamma-ray or electron), 3) all SiPM pixels have a zero dark count rate and avalanche triggering probability, and 4) there is no Philips DPC3200 SiPM onboard sub-pixel or validation trigger logic. Finally, the output of the Philips DPC3200 SiPM per simulate primary particle was implemented to approximate the unit output: an 8$\times$8 array of representing the total number of SPAD triggers per SiPM pixel. However, to enable further analysis to optimise the PET detector design, each 8$\times$8 SiPM pixel SPAD trigger count was also accompanied by a full list their respective timestamps relative to the first interaction time of the primary particle within the LYSO crystals.

\begin{figure}[tbh]   
   \centering
   \includegraphics[width=0.4\textwidth]{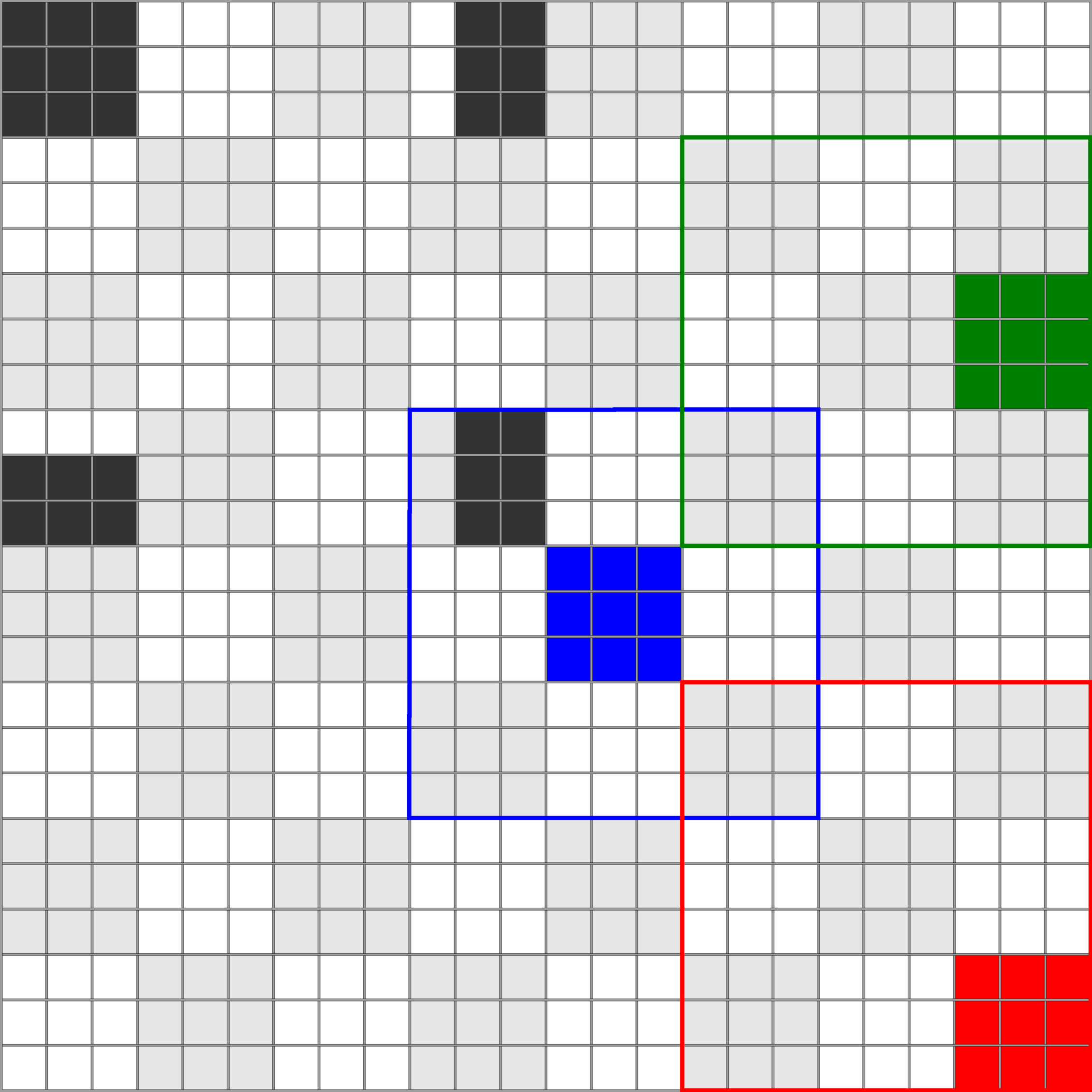}
\caption{The relative position of 3$\times$3 SiPM pixel footprints for maximum SiPM pixel values in the central (blue), edge (green) and corner (red) regions within the 8$\times$8 photosensor SiPM array. Here the solid coloured blocks correspond respectively to each of the maximum SiPM pixel value locations. In addition, the twenty seven different LYSO crystal locations that capitalised on system symmetry to calculate the RM is shown in black shading.}
\label{fig:5}
\end{figure}

Interaction position of each simulated gamma-ray within the PET radiation detector LYSO scintillator crystal array was determined though the use of a Weighted Least Square (WLS) algorithm \cite{Ling2008}. The WLS algorithm utilises an minimisation approach where the photosensor response model output for an event of interest, or also known as a Data Measurement ($\text{DM}$), is compared to an array of photosensor response model outputs corresponding to known gamma-ray/gamma-ray surrogate interaction locations (known as Reference Measurements ($\text{RM}$)). This approach of estimating gamma-ray interaction position $(x,y,z)$ can be expressed as:

\begin{equation}
    \displaystyle \left(x,y,z,RM\right) = \genfrac{}{}{0pt}{2}{\displaystyle {\text{argmin}}}{\left(x,y,z,RM\right)}\sum^{9}_{i=1}\frac{\left(\text{DM}-\text{RM}\right)^2}{\text{RM}}
\end{equation}

\noindent where both the $\text{DM}$ and $\text{RM}$ arrays are limited to a 3$\times$3 SiPM pixel footprint to match the corresponding expected light sharing range of the encoded ESR foil array pattern (i.e. $\pm$ 3 crystal). The orientation of this 3$\times$3 SiPM pixel footprint limitation within the 8$\times$8 photosensor SiPM array is determined by the location of maximum SiPM pixel value. Figure \ref{fig:5} illustrates how this 3$\times$3 SiPM pixel footprint was orientated for a maximum SiPM pixel value in the central (blue), edge (green), and corner (red) regions of the photosensor SiPM pixel array.

The RM contains a set of 14 surrogate depth dependent photoelectrically absorbed 511 keV gamma-ray 3$\times$3 SiPM pixel footprints for each individual LYSO crystal within the encoded Vikuiti ESR foil separated and wrapped array. These surrogate 511 keV gamma-ray interaction depth dependent 3$\times$3 SiPM pixel footprints were calculated, on a 1 mm resolution along the depth of the PET detector (Z direction), with the developed Geant4 test platform for five hundred electrons emitted in a 2$\pi$ solid angle at the center of x-y cross-section of a select number of crystal LYSO. Twenty seven different LYSO crystal locations, seen in Figure \ref{fig:5}, were selected to capitalise on the PET detectors symmetry and the individual pixel mean 3$\times$3 SiPM footprints calculated to populate the RM. 

\subsubsection{PET Radiation Detector Performance Assessment/Optimisation}
\label{sec:m1o}

\begin{figure}[tbh]   
   \centering
   \includegraphics[width=0.4\textwidth]{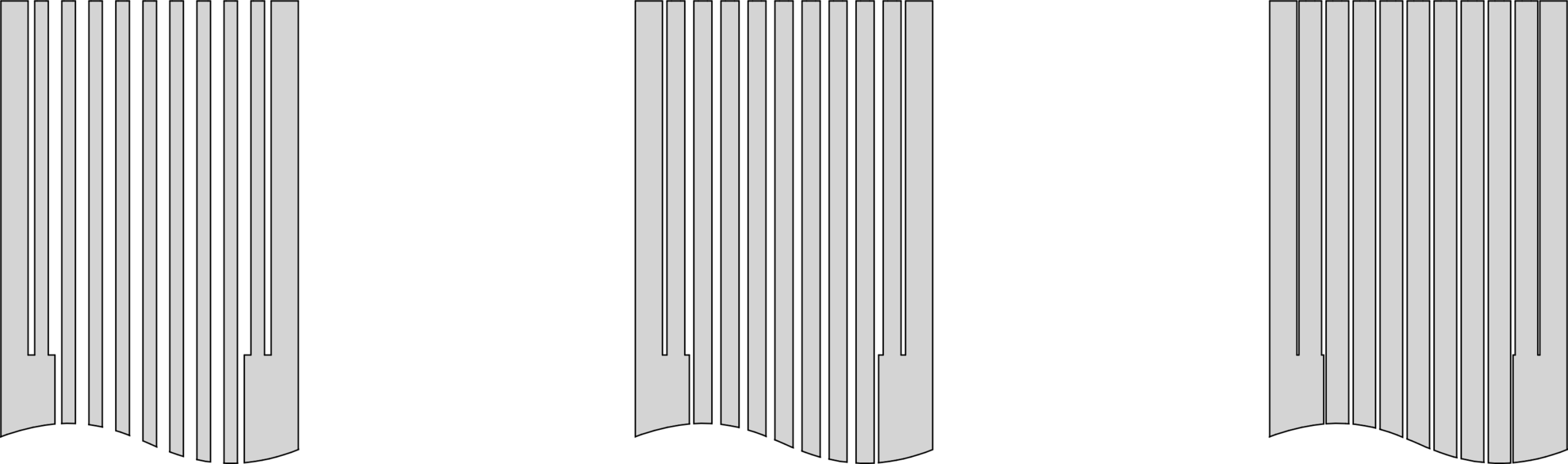}
\caption{Top tenth of central PSI filled 5 step encoded reflective foil design seen in Figure \ref{fig:1} with PSI dilation values of 1.5, 2.0 and 2.5. A reduction in open cross-section for propagation of optical photons between LYSO crystal can be observed.}
\label{fig:6}
\end{figure}

The impact of two physical properties were explored to assess/optimise the performance of the proposed PET radiation detector: LYSO crystal surface roughness, and encoded reflective foil PSI width. Three different surface roughnesses of 0.1, 2.8 and 5.6 degrees were simulated to approximate optical surface properties of polished, ground and cut LYSO crystals \cite{Moisan1997}. Whereas for the encoded reflective foil PSI width, thirteen different PSI width dilation's over a range of 1 to 2.5 in steps of 0.125 were simulated. Here a PSI dilation value of 1 was set to be the default seen in Figure \ref{fig:1}, with comparative examples of its impact on encoded reflective foil structure for the values of 1.5, 2 and 2.5 seen in Figure \ref{fig:6}. 

For each combination of surface roughness and PSI dilation, a total of 250,000 511 keV gamma-rays was simulated from a point source 350 mm away in front of the LYSO crystal array and limited in angular emission towards the arrays top outer edges. Assessment of the performance of the PET radiation detector in these configurations was determine through the use of five Figures of Merit (FoMs): 511 keV photopeak Full Width at Half Maximum (FWHM) energy resolution, Crystal of Interaction Identification Accuracy (CoIIA), estimated DoI accuracy, extent of Light Restriction (LR) to a 3$\times$3 SiPM pixel footprint, and relative SPAD trigger time for the 1st, 10th and last optical photon per 511 keV gamma-ray. The last four of these five FoMs was applied to photoelectric absorption events of the incident 511 keV gamma-rays, with all FoMs calculated for four LYSO crystal array region classifications: central, edge (within 3 crystal of a single array edge), corner (within 3 crystal of two array edges), and total.

\section{Results}
\label{sec:R}
\begin{figure}[tbh]    
    \centering
    \begin{subfigure}
    \centering 
        \includegraphics[width=0.235\textwidth, trim = {5 0 30 10},clip]{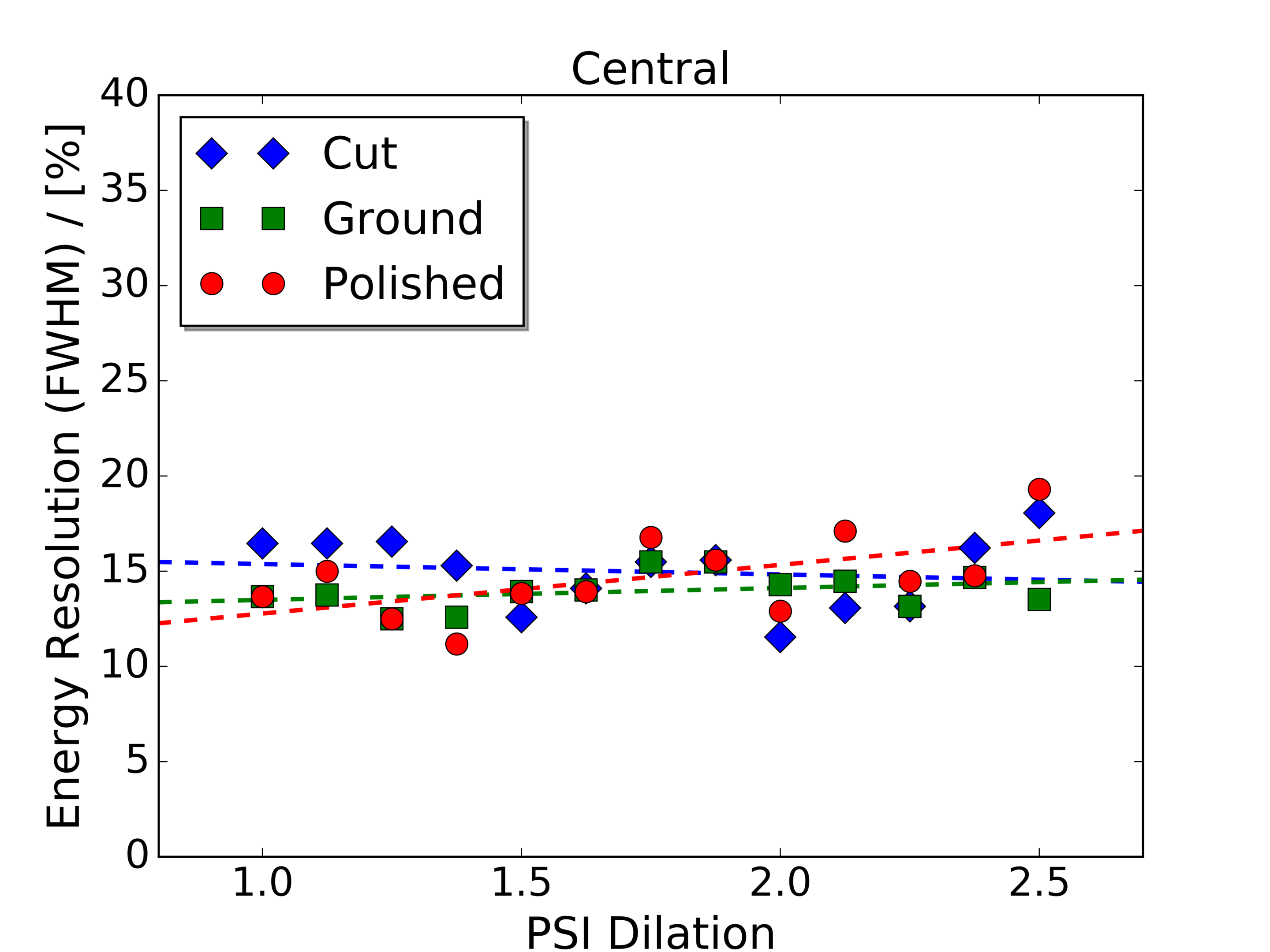}
        \label{fig:r1a}
    \end{subfigure}
    \begin{subfigure}
    \centering
        \includegraphics[width=0.235\textwidth, trim = {5 0 30 10},clip]{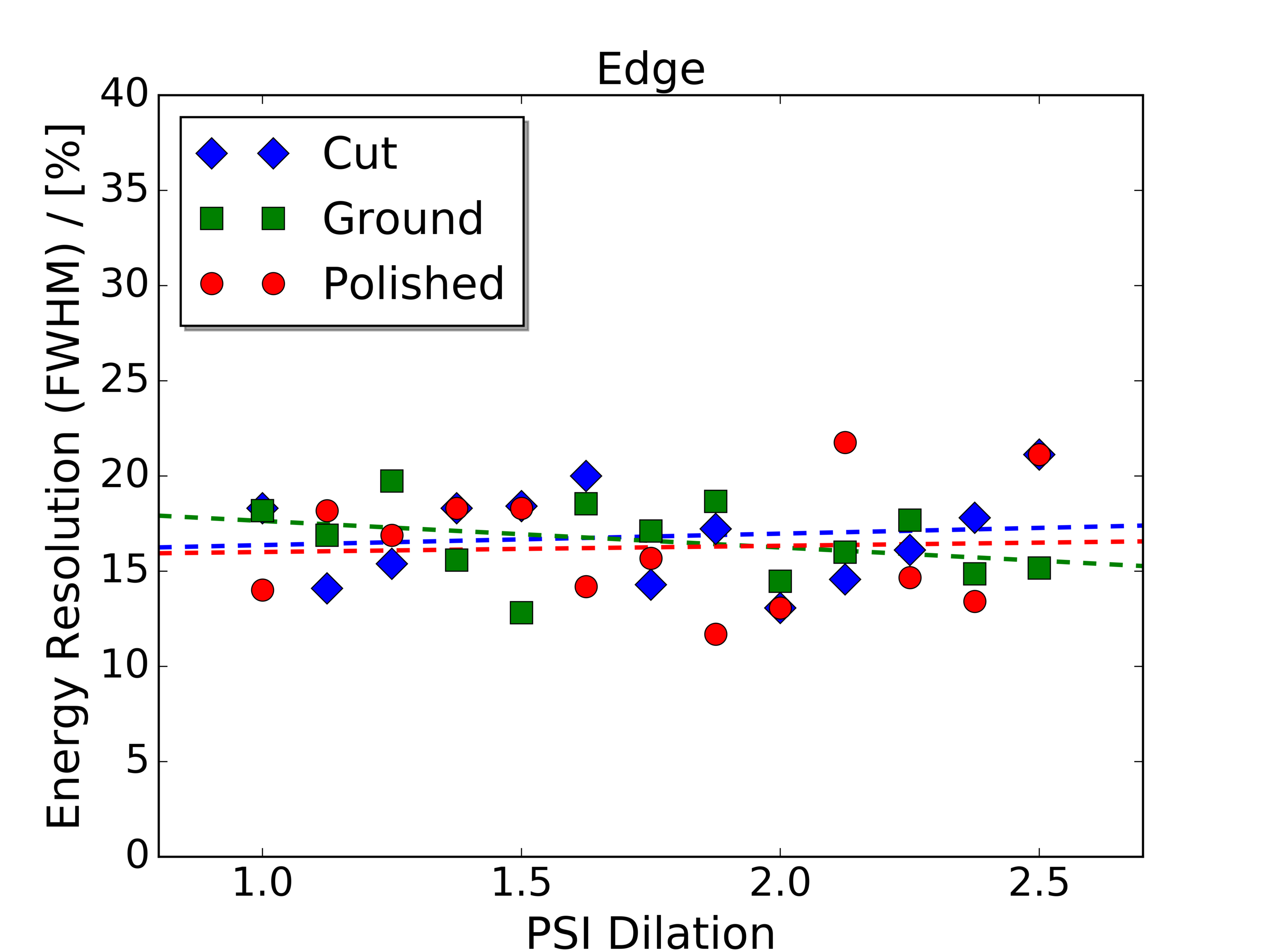}
        \label{fig:r1b}
    \end{subfigure}
    \newline
    \begin{subfigure}
    \centering 
        \includegraphics[width=0.235\textwidth, trim = {5 0 30 10},clip]{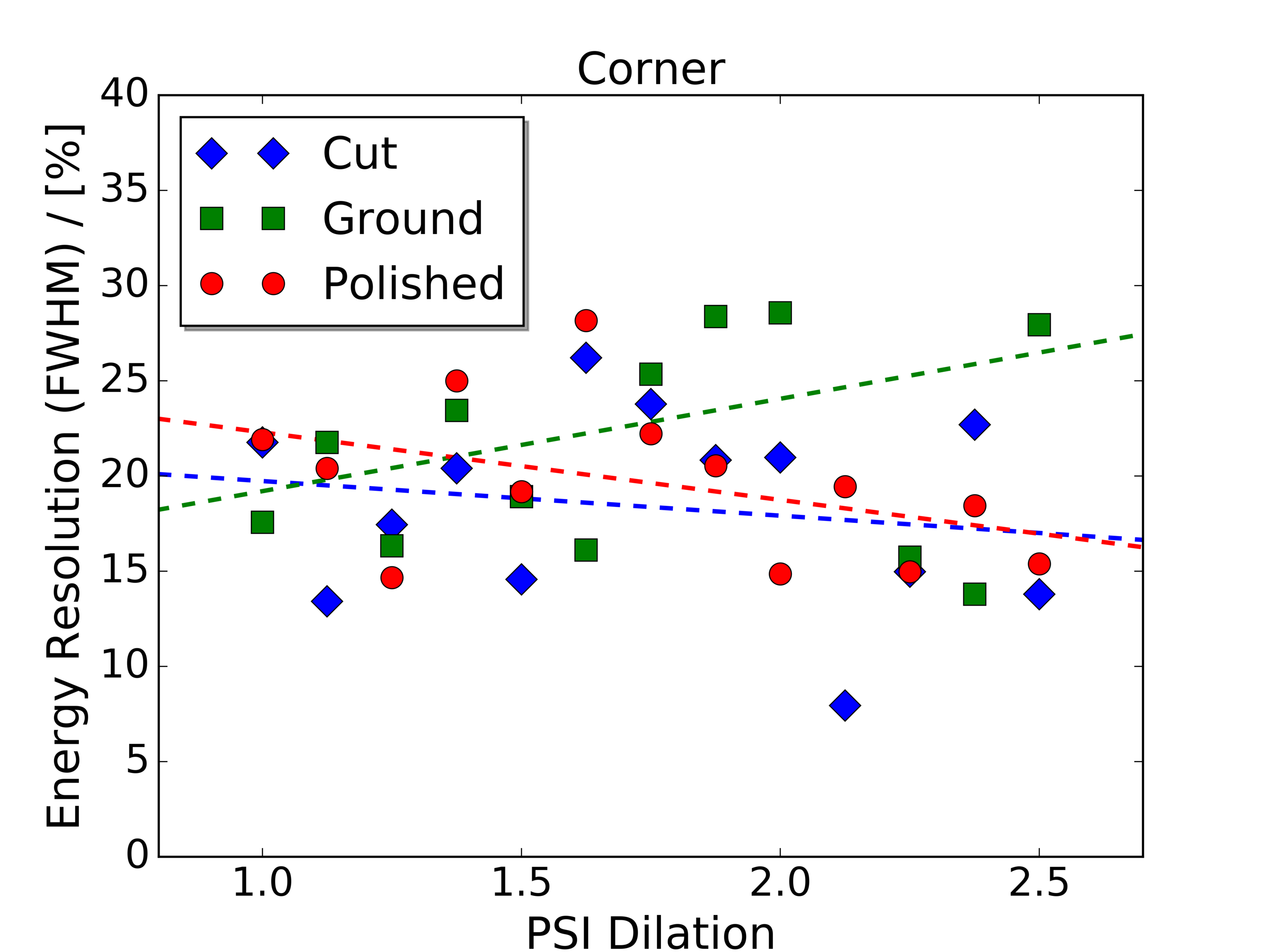}
        \label{fig:r1c}
    \end{subfigure}
    \begin{subfigure}
    \centering
        \includegraphics[width=0.235\textwidth, trim = {5 0 30 10},clip]{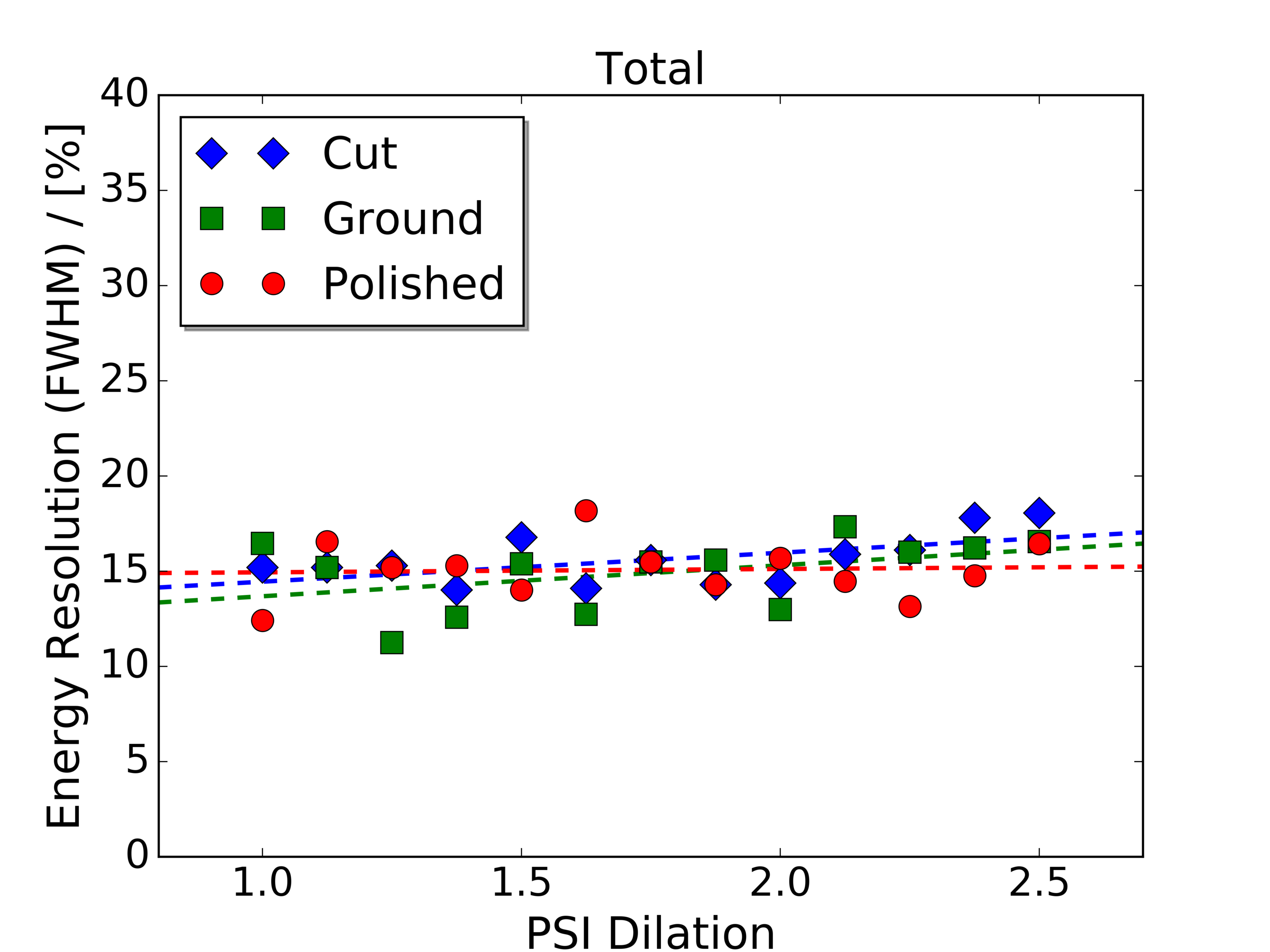}
        \label{fig:r1d}
    \end{subfigure}
\caption{Energy resolution (FWHM) for four LYSO array crystal region classifications: central (top left), edge (top right), corner (bottom left), and total (bottom right). The coloured dash lines correspond to a fitted linear function for each crystal surface type to illustrate the general trend as a function of PSI.}
\label{fig:r1}
\end{figure}

The 511 keV photopeak FWHM energy resolution of the three different crystal surface types and four different LYSO crystal array region classifications as a function of PSI dilation can be seen in Figure \ref{fig:r1}. As is typically observed in crystal array based PET radiation detectors, the effective energy resolution in the central region of the array is generally superiour to the edge and corner regions for all crystal surface type and PSI combinations \cite{Cherry2003,Bushberg2011}. Furthermore, the crystal surface roughness and PSI dilation seem to have minimal impact on energy resolution. This indicates that the potential total signal loss due to SPAD saturation from the restriction of light sharing range, i.e. multiple optical photons striking the same SPAD per gamma-ray, outweighs the impact of light trapping between crystals because of the presence of the PSIs. Therefore, based on this data, an energy resolution of approximately 15\% would be expected regardless of the selected crystal surface conditions and PSI dilation.

\begin{figure}[tbh]    
    \centering
    \begin{subfigure}
    \centering 
        \includegraphics[width=0.235\textwidth, trim = {5 0 30 10},clip]{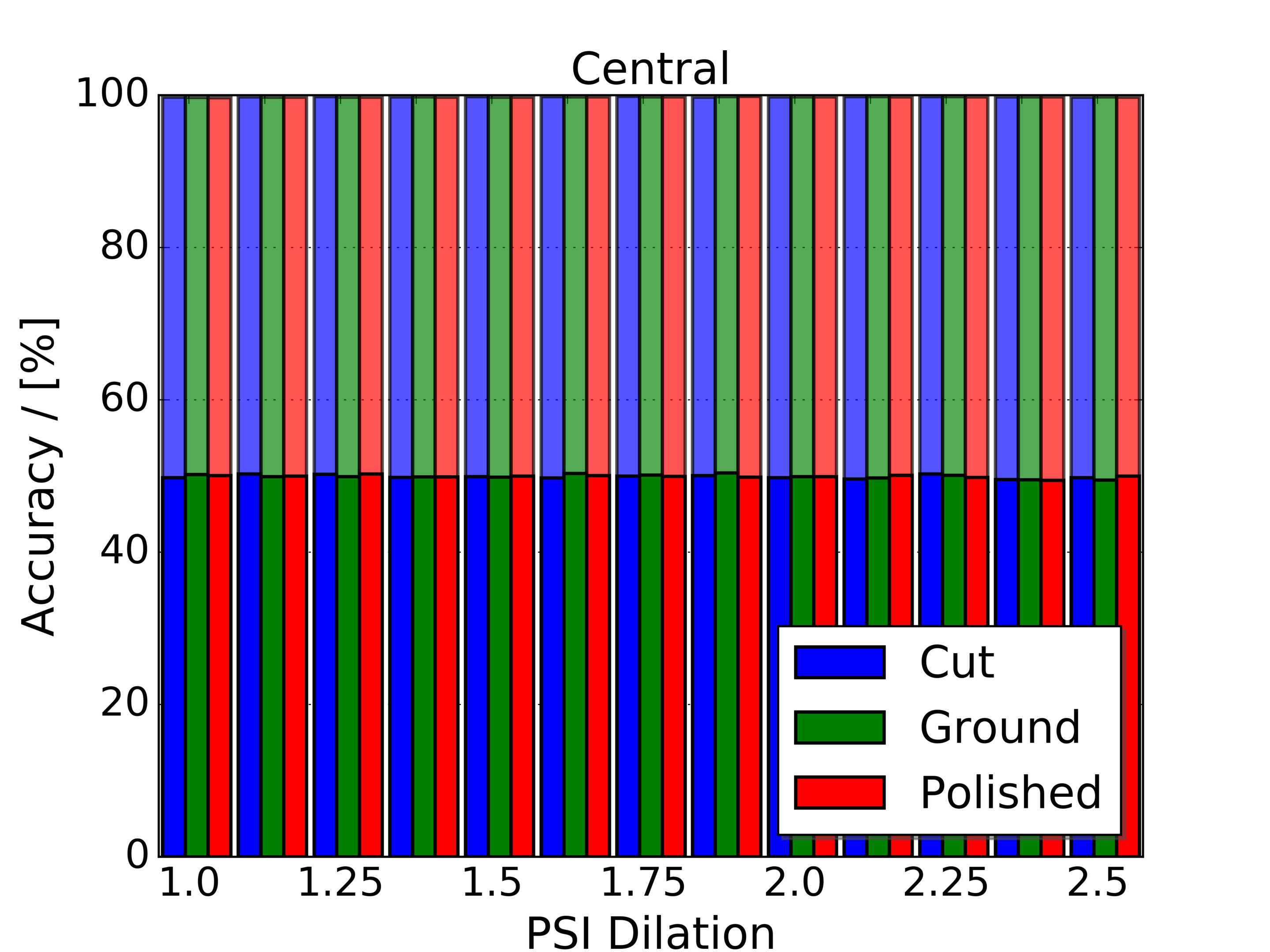}
        \label{fig:r2a}
    \end{subfigure}
    \begin{subfigure}
    \centering
        \includegraphics[width=0.235\textwidth, trim = {5 0 30 10},clip]{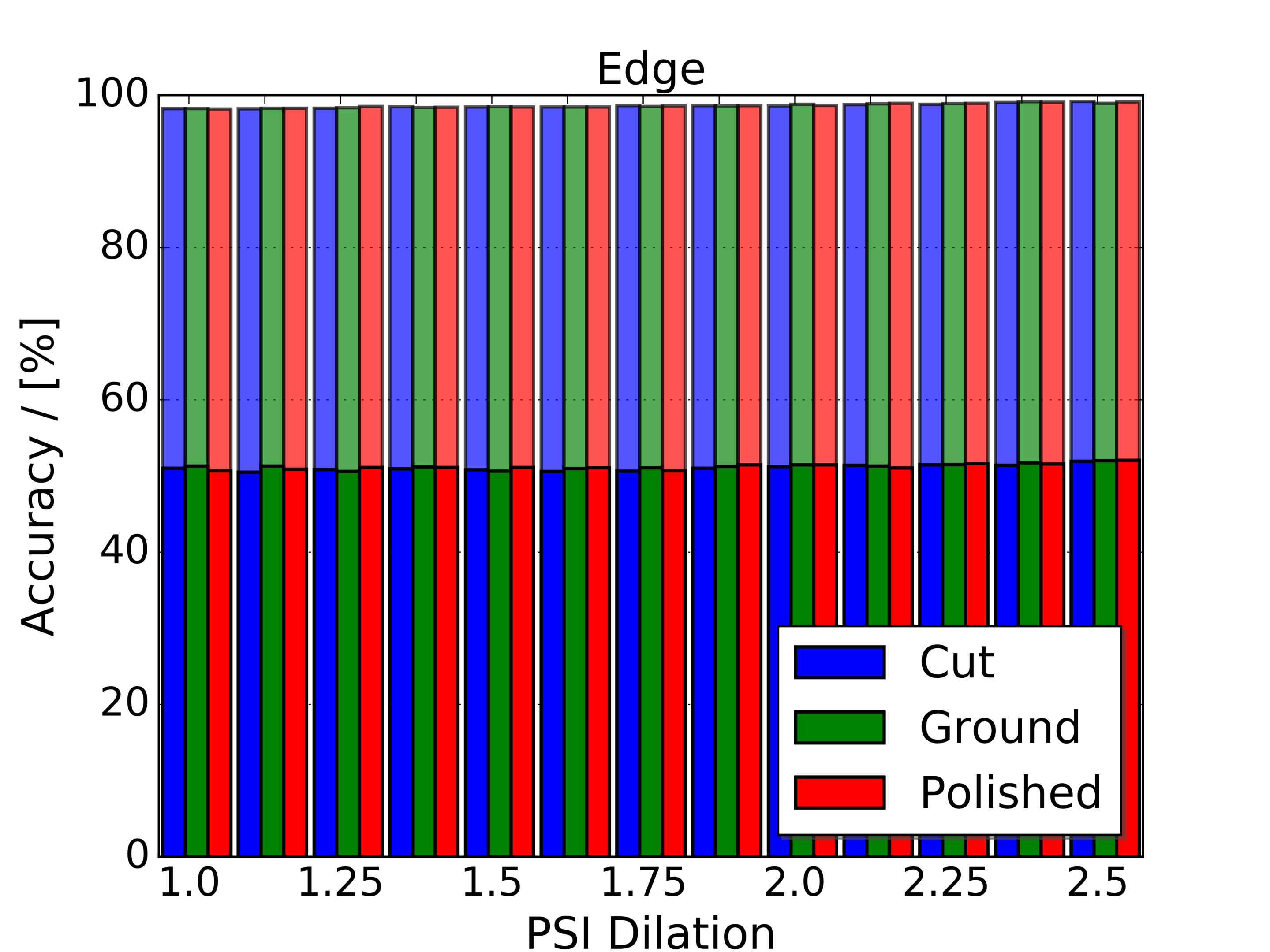}
        \label{fig:r2b}
    \end{subfigure}
    \newline
    \begin{subfigure}
    \centering 
        \includegraphics[width=0.235\textwidth, trim = {5 0 30 10},clip]{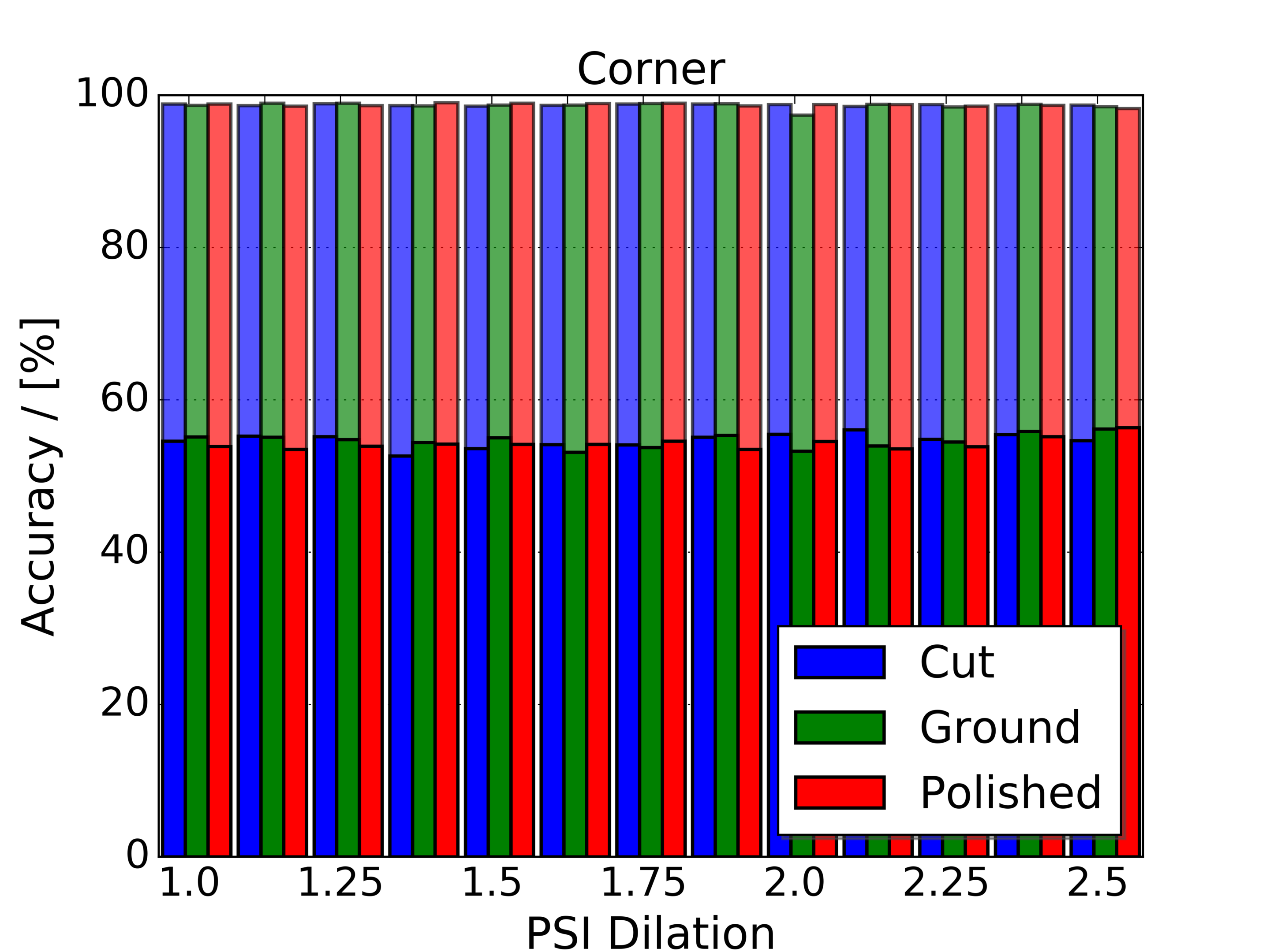}
        \label{fig:r2c}
    \end{subfigure}
    \begin{subfigure}
    \centering
        \includegraphics[width=0.235\textwidth, trim = {5 0 30 10},clip]{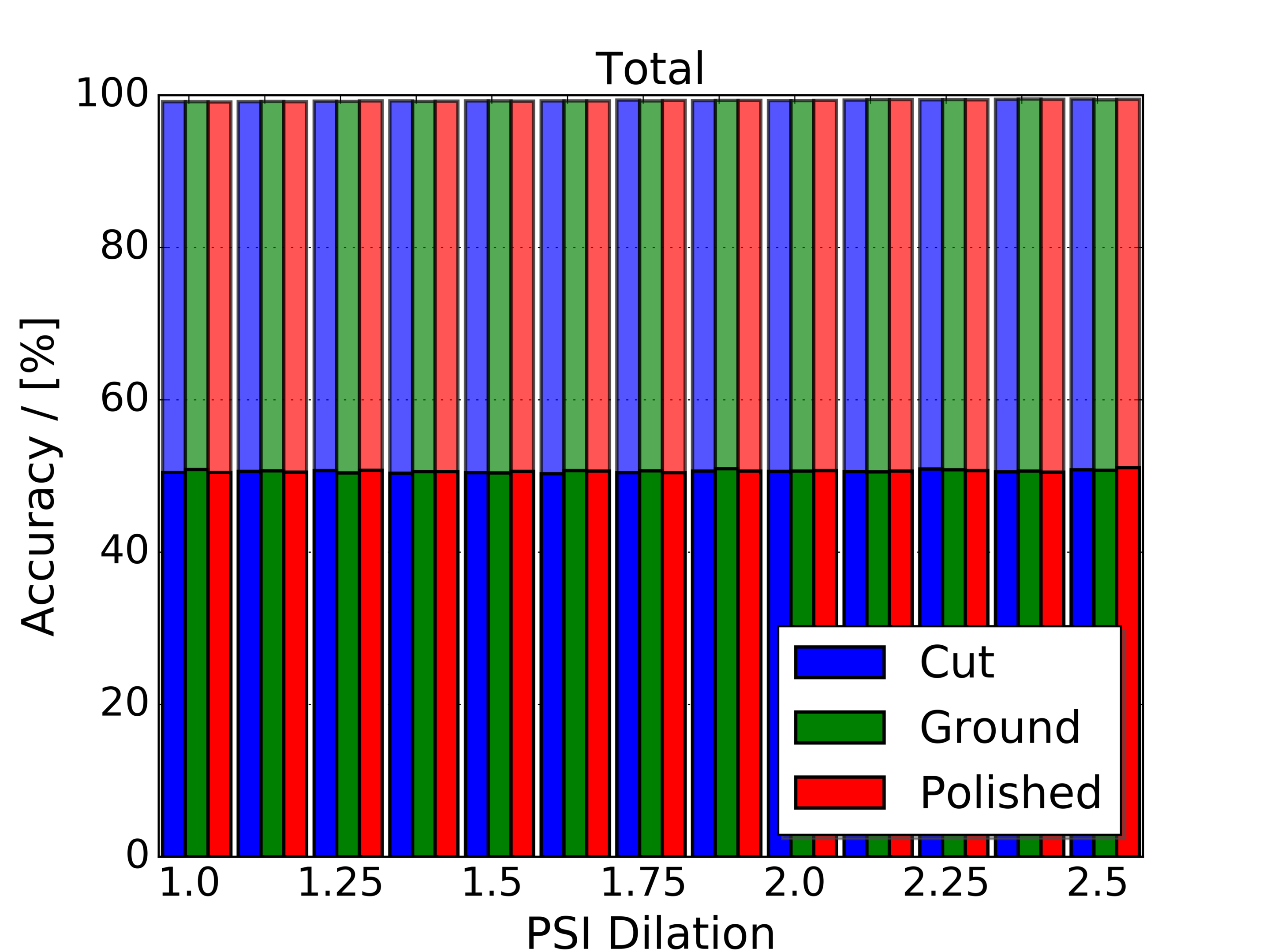}
        \label{fig:r2d}
    \end{subfigure}
\caption{CoIIA for four LYSO array crystal region classifications: central (top left), edge (top right), corner (bottom left), and total (bottom right). For each PSI dilation value three sets of bars can be seen that correspond, from left to right, to cut, ground, and polished crystal surface roughness data. The two shades of CoIIA data, in decreasing intensity, represent the accuracy of estimating gamma-ray interaction within the ``true" crystal of interaction and its neighbour.}
\label{fig:r2}
\end{figure}

Figure \ref{fig:r2} presents the CoIIA of the three different crystal surface types and four different LYSO crystal array region classifications as a function of PSI dilation. In contrast to the effective energy resolution trends observed in Figure \ref{fig:r1}, the edge and corner regions within the LYSO crystal array possess higher CoIIA than the central region. When the range of CoIIA is expanded to include the neighbouring pixels as well, this relationship reverts to match the general behaviour that the performance of the central crystal array region is superiour. Furthermore, this expanded x-y crystal range ($\pm$ 1 crystal) also results in the CoIIA approaching 100\% for all crystal regions. However regardless of CoIIA range their appears to be a near zero effect due to either the PSI dilation or crystal surface type, with the ``true” crystal of interaction always begin identified over 50\% of the time. 

\begin{figure}[tbh]    
    \centering
    \begin{subfigure}
    \centering 
        \includegraphics[width=0.235\textwidth, trim = {5 0 30 10},clip]{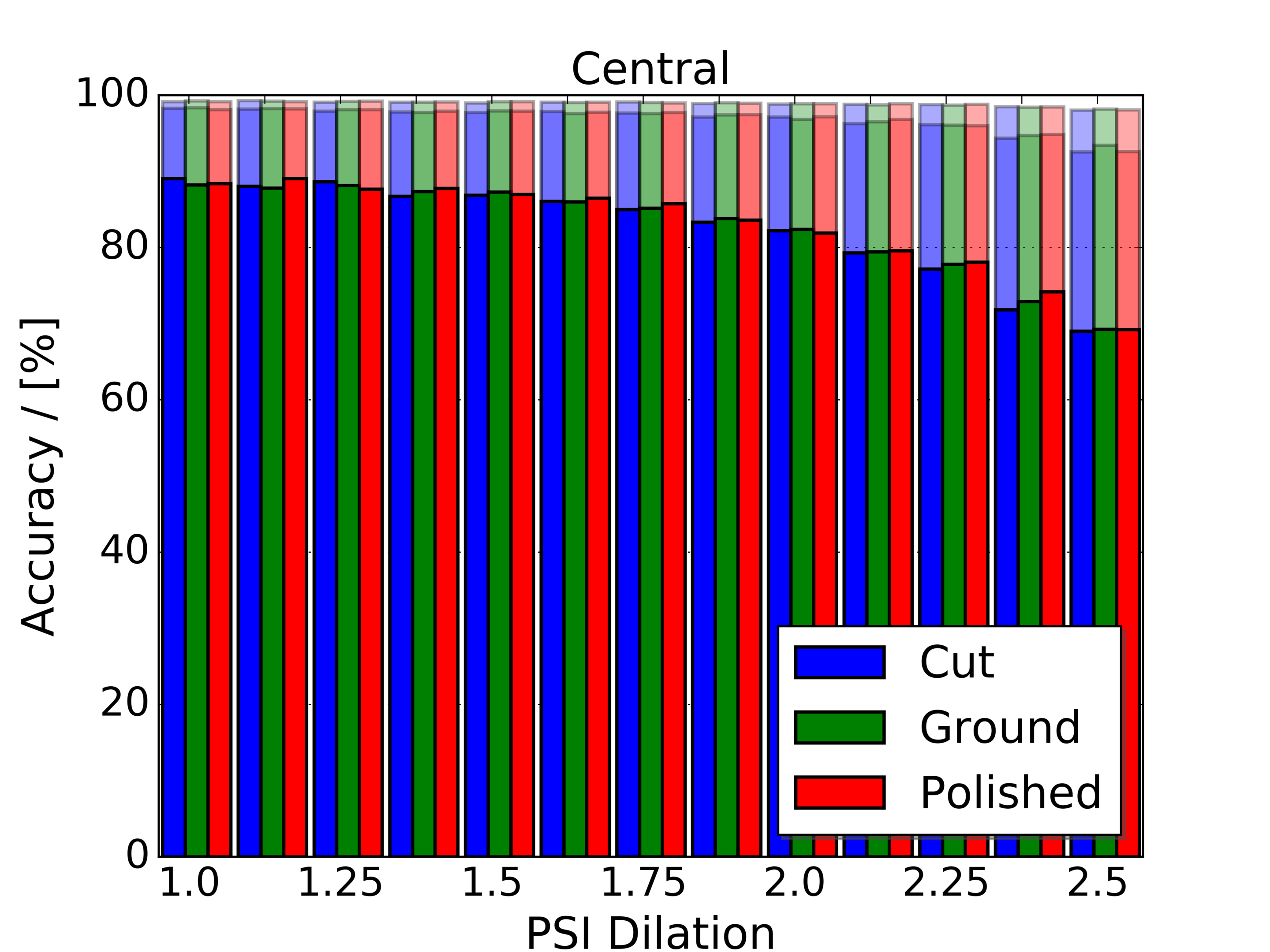}
        \label{fig:r3a}
    \end{subfigure}
    \begin{subfigure}
    \centering
        \includegraphics[width=0.235\textwidth, trim = {5 0 30 10},clip]{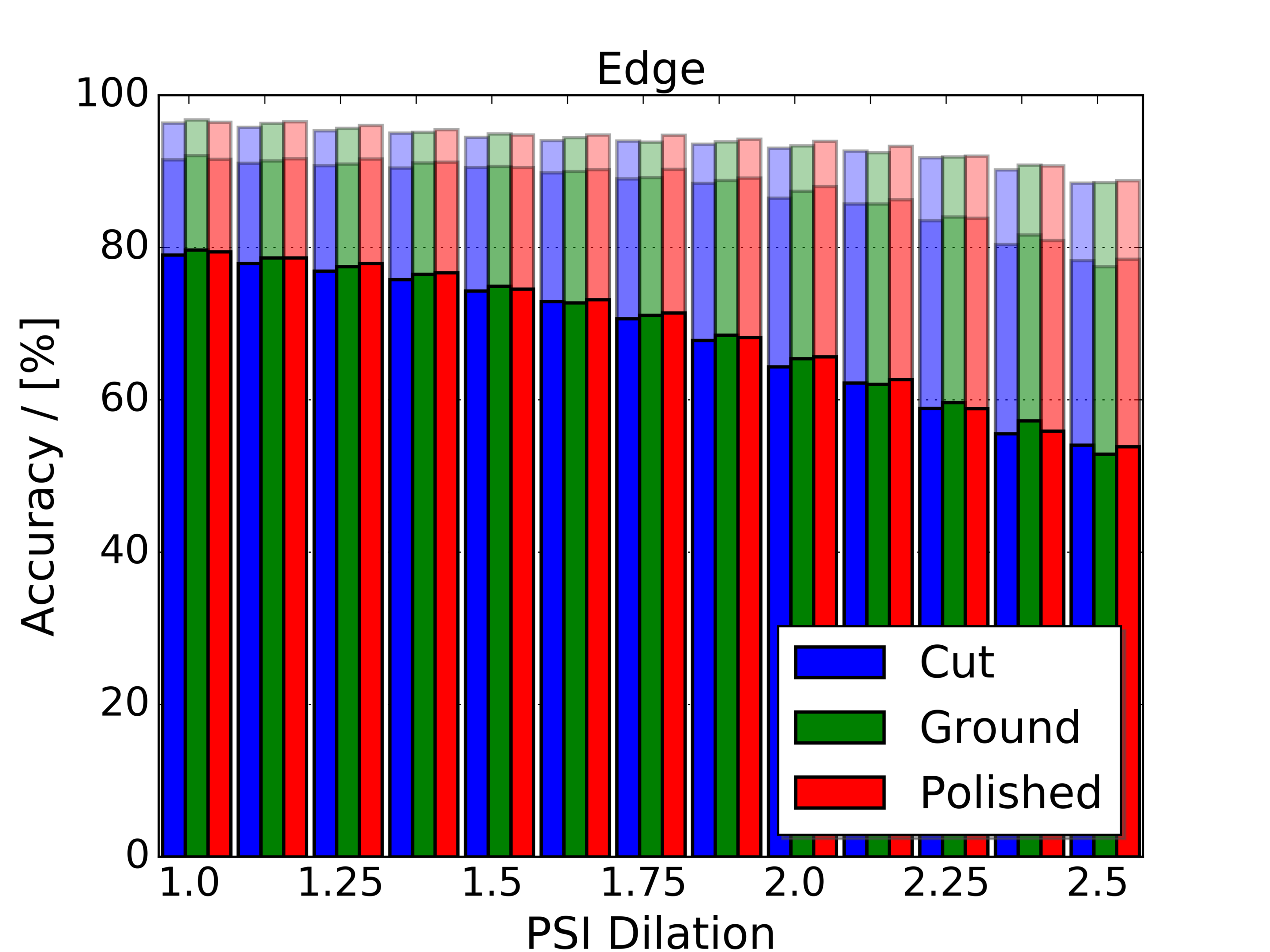}
        \label{fig:r3b}
    \end{subfigure}
    \newline
    \begin{subfigure}
    \centering 
        \includegraphics[width=0.235\textwidth, trim = {5 0 30 10},clip]{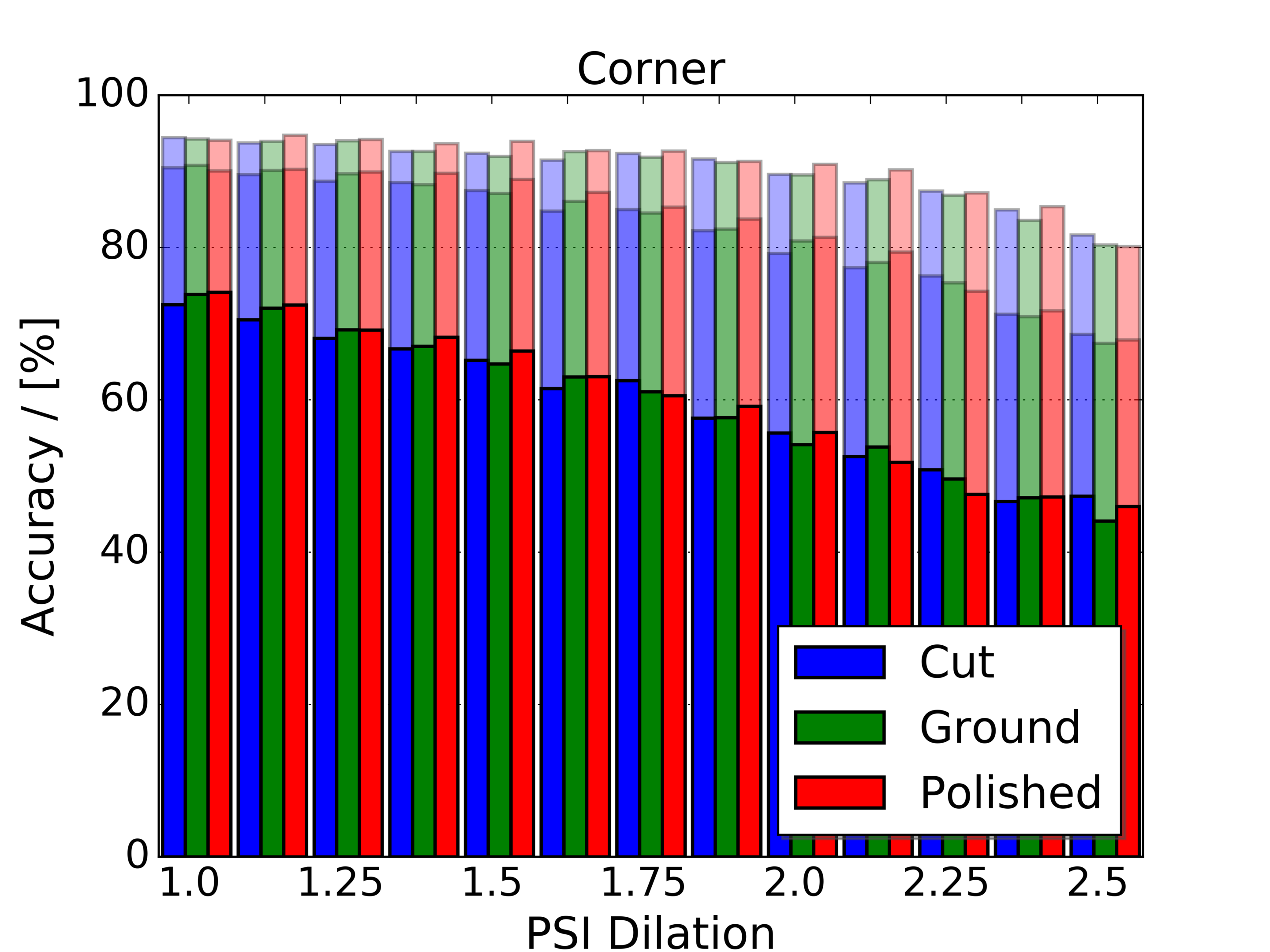}
        \label{fig:r3c}
    \end{subfigure}
    \begin{subfigure}
    \centering
        \includegraphics[width=0.235\textwidth, trim = {5 0 30 10},clip]{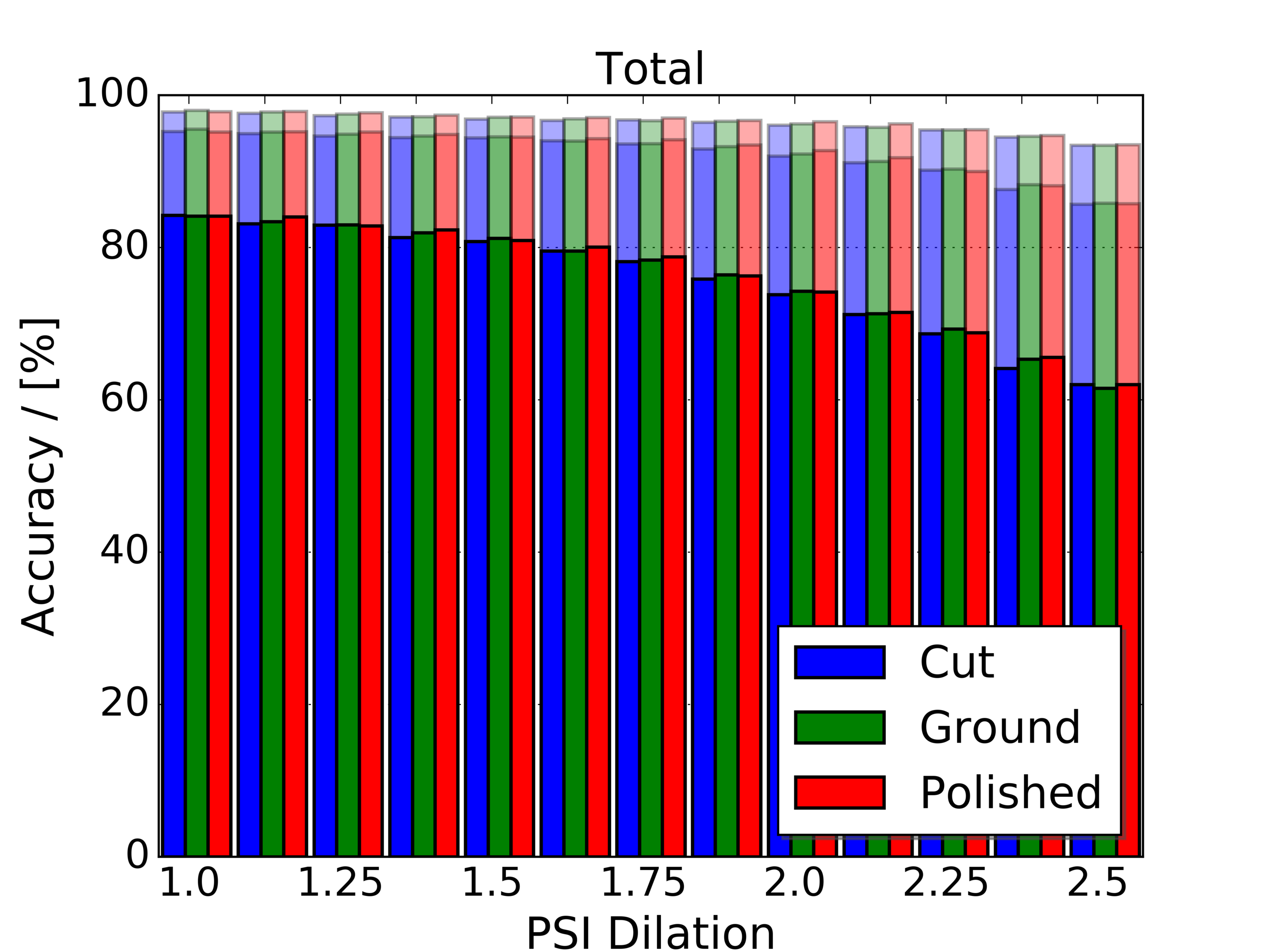}
        \label{fig:r3d}
    \end{subfigure}
\caption{Estimated DoI for four LYSO array crystal region classifications: central (top left), edge (top right), corner (bottom left), and total (bottom right). For each PSI dilation value three sets of bars can be seen that correspond, from left to right, to cut, ground, and polished crystal surface roughness data. The decreasing intensity of data shading corresponds to the accuracy of estimating the gamma-ray interaction to within 2, 4 and 6 mm respectively.}
\label{fig:r3}
\end{figure}

\begin{figure}[tbh]    
    \centering
    \begin{subfigure}
    \centering 
        \includegraphics[width=0.235\textwidth, trim = {5 0 30 10},clip]{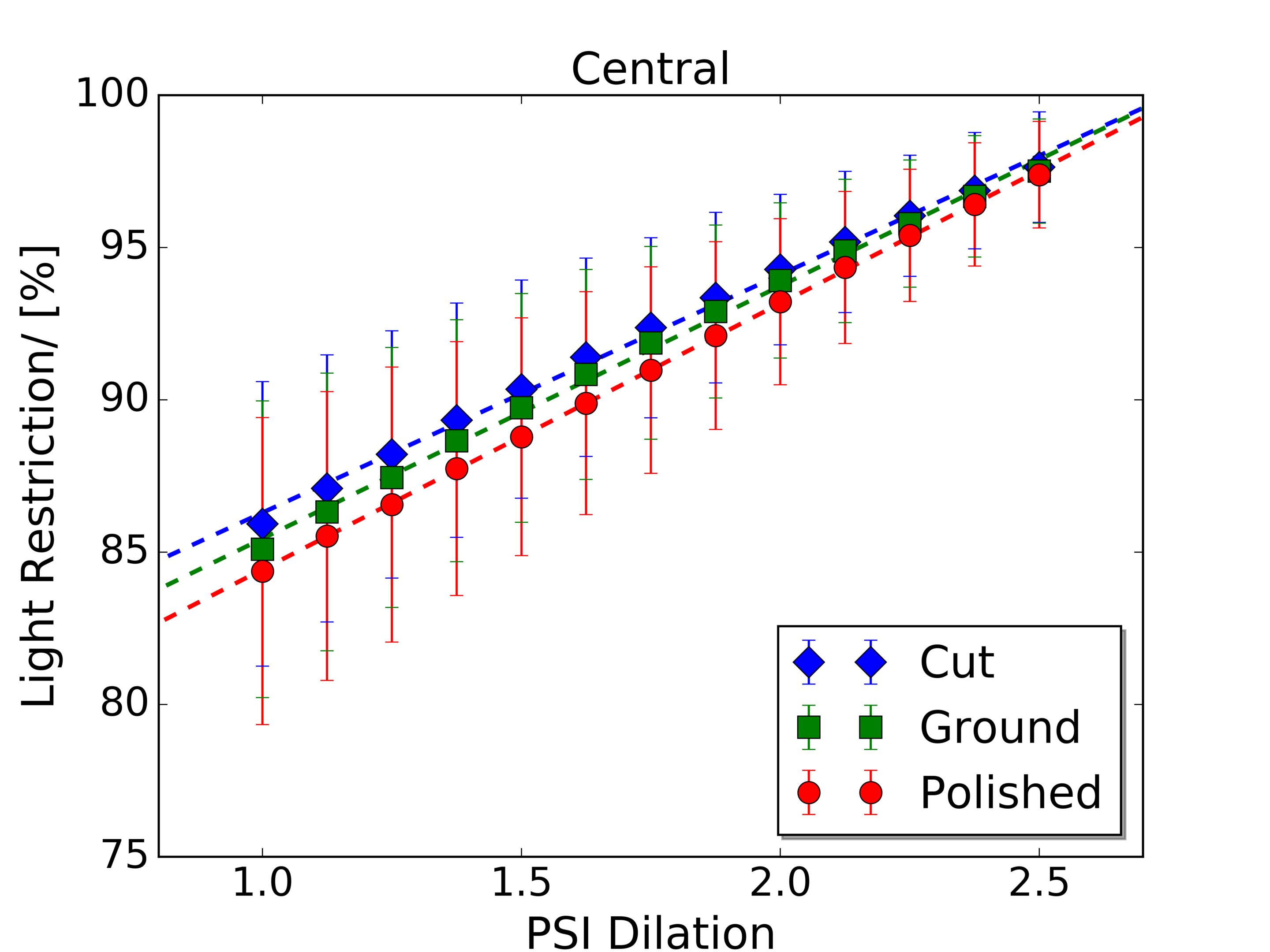}
        \label{fig:r4a}
    \end{subfigure}
    \begin{subfigure}
    \centering
        \includegraphics[width=0.235\textwidth, trim = {5 0 30 10},clip]{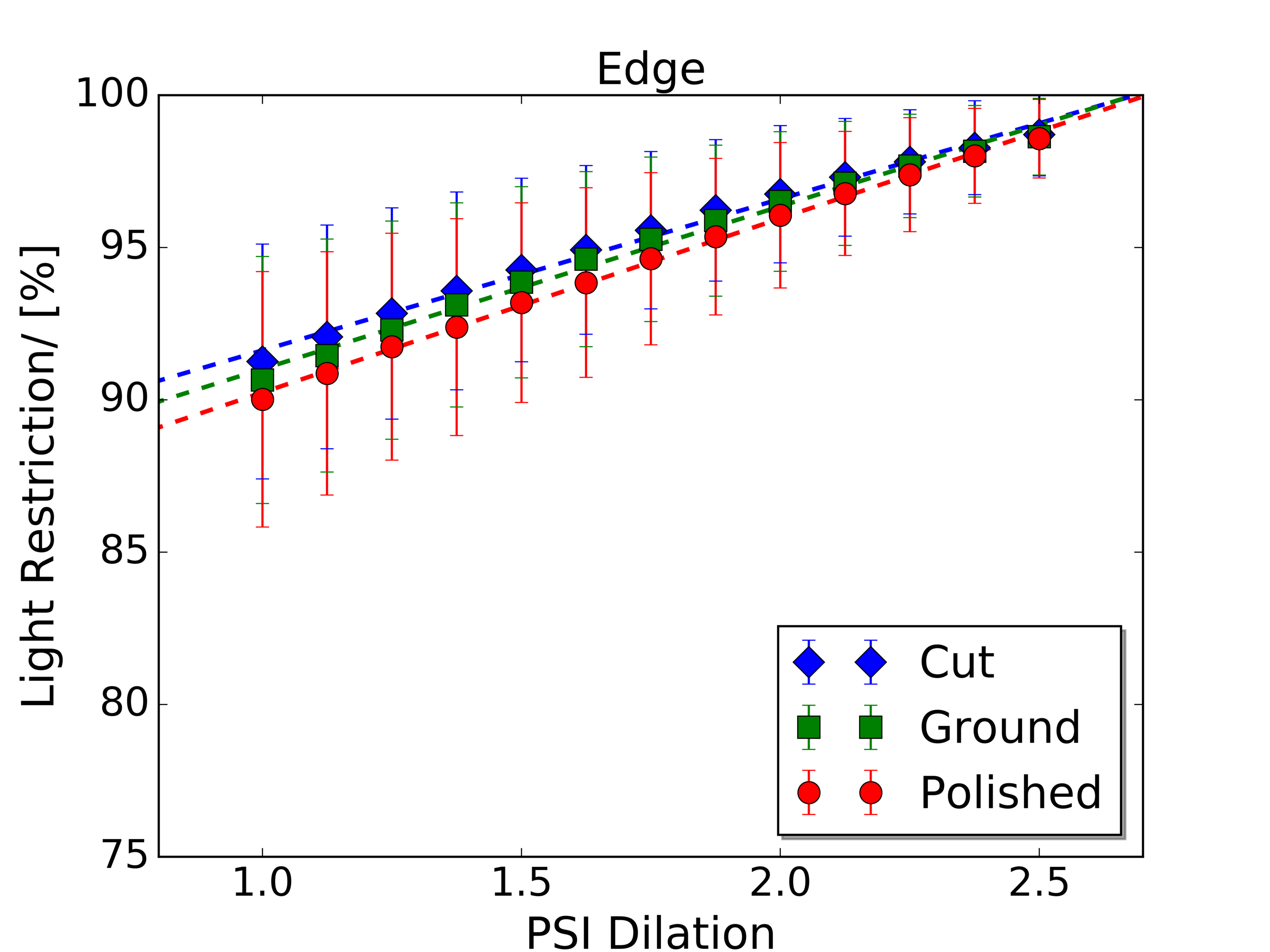}
        \label{fig:r4b}
    \end{subfigure}
    \newline
    \begin{subfigure}
    \centering 
        \includegraphics[width=0.235\textwidth, trim = {5 0 30 10},clip]{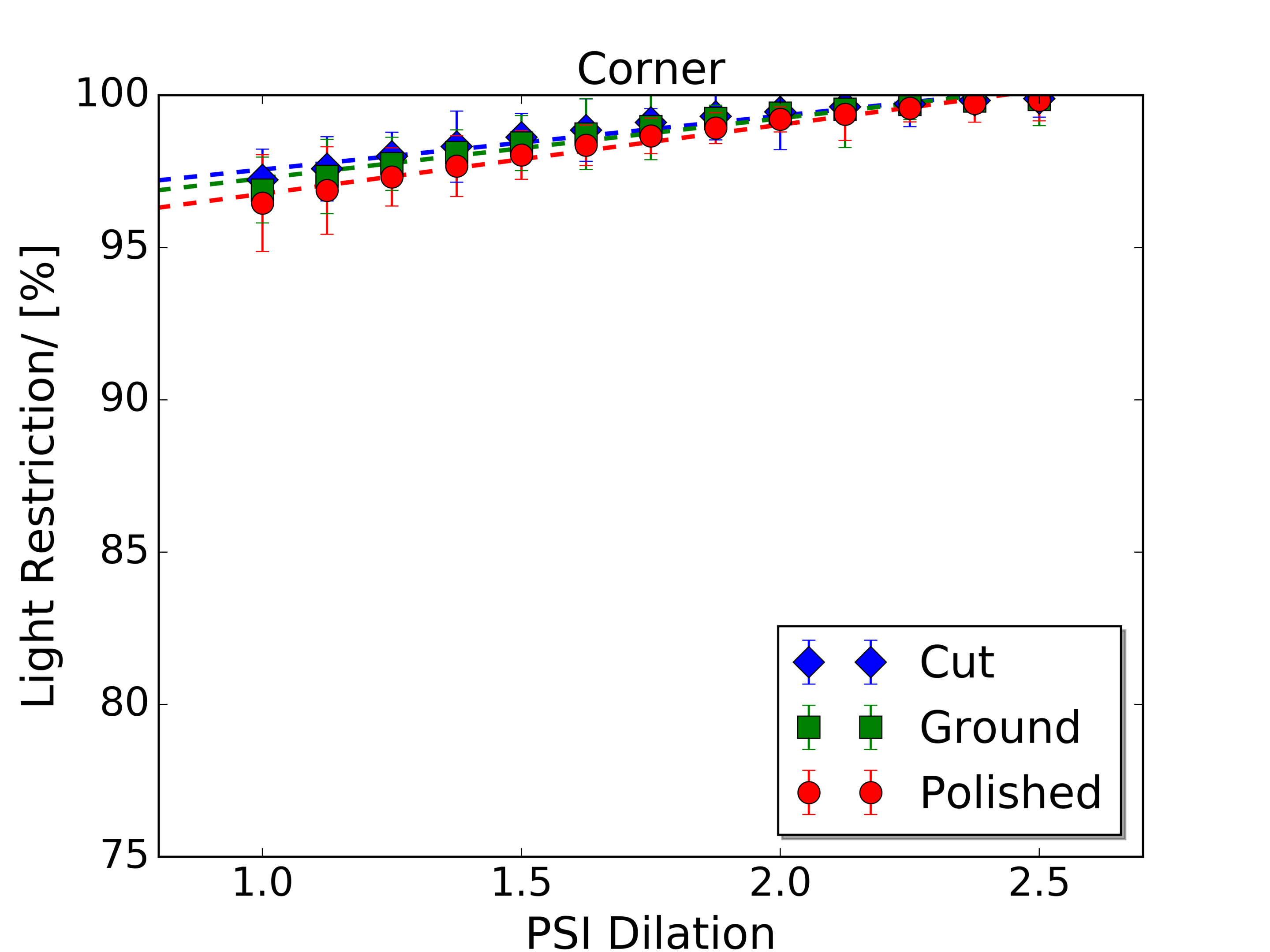}
        \label{fig:r4c}
    \end{subfigure}
    \begin{subfigure}
    \centering
        \includegraphics[width=0.235\textwidth, trim = {5 0 30 10},clip]{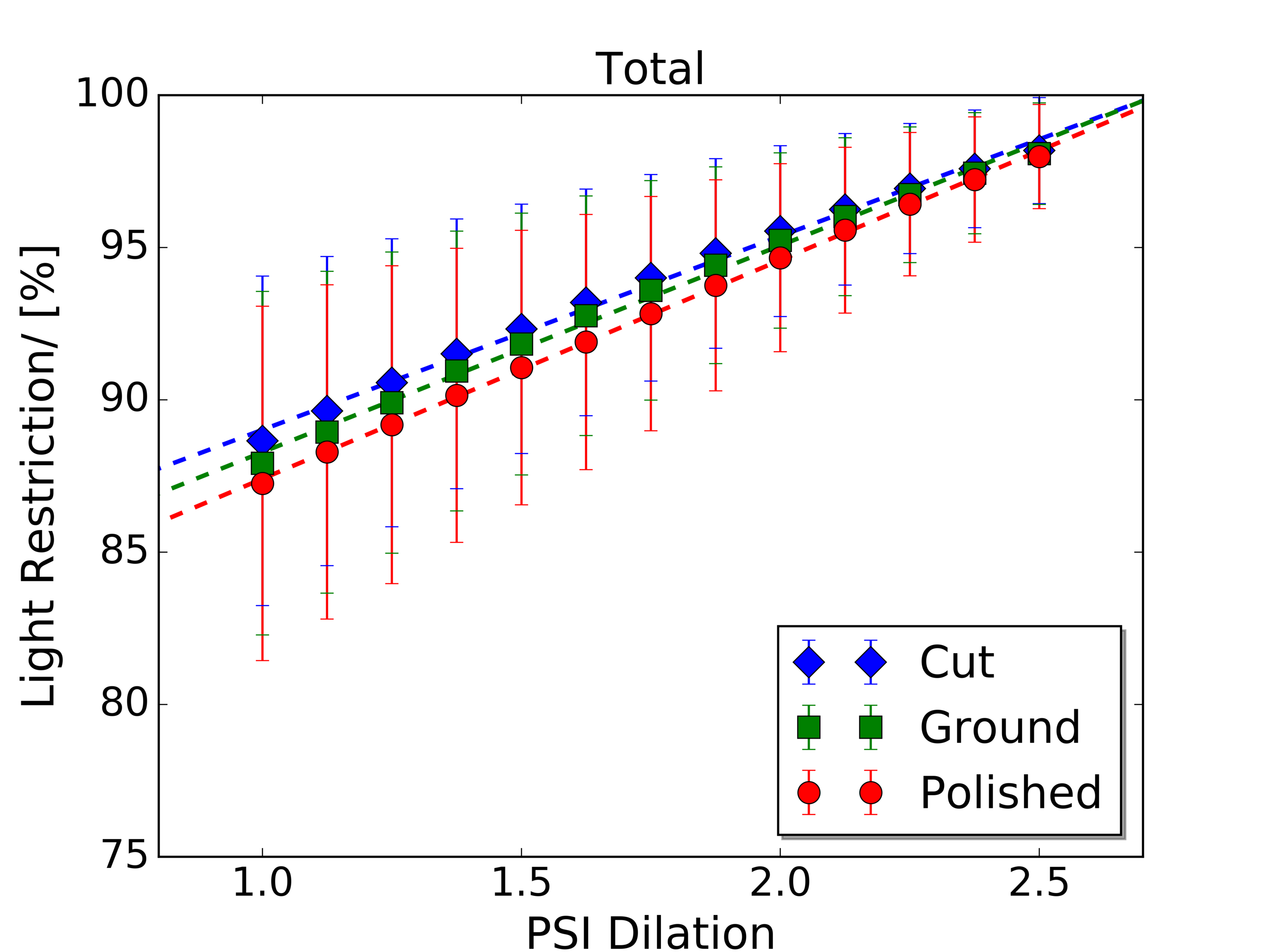}
        \label{fig:r4d}
    \end{subfigure}
\caption{Mean (markers) and standard deviations (bounding bars) of LR to a 3$\times$3 SiPM pixel footprint for four LYSO array crystal region classifications: central (top left), edge (top right), corner (bottom left), and total (bottom right). The coloured dash lines correspond to a fitted linear function for each crystal surface type to illustrate general trends as a function of PSI.}
\label{fig:r4}
\end{figure}

The estimated DoI accuracy to within 2, 4 and 6 mms of actual gamma-ray interaction depth is presented in Figure \ref{fig:r3}. This figure shows again that the central crystal array region possess superiour performance over the edge and corner regions \cite{Cherry2003,Bushberg2011}. Across the PSI dilation range the DoI estimation accuracy to within 2 mm can be seen to be 10 and 20 \% lower for the edge and corner regions respectively, regardless of the LYSO crystal surface type. In the case of the 4 mm and 6 mm data the observed difference is less, but still present. However, in contrast to the previously discussed FoMs, clear dependencies of DoI performance can be observed for both the crystal surface type and PSI dilation. In the case of the crystal surface it appears that an inverse relationship exists between surface roughness and DoI performance (i.e. a polished crystal surface would yield the best DoI performance for the proposed PET radiation detector design). This is most likely due to the higher probability that optical photons will propagate between crystal with minimal scattering for smoother surfaces, ensuring a higher faction make it to the second, or even third, crystal away from the crystal of origin (support for this effect can seen within the LR FoM results seen in Figure \ref{fig:r4}). Whereas for PSI dilation, a clear inverse relationship with DoI accuracy is present regardless of crystal array region classification. 

For the extent of LR as a function of gamma-ray interaction position with in the crystal array, seen in Figure \ref{fig:r4}, three of notable trends as a function of PSI dilation and crystal surface roughness can be observed. The first of these trends is that the extent of LR is directly proportional to crystal surface roughness (i.e. high surface roughness leads to greater internal light scattering within each LYSO crystal). Second, a direct relationship between LR and PSI dilation is present due to the reduction in total open cross-section of the foils limiting light propagation between LYSO crystals. Third, at the edge and corner regions within the LYSO crystal array the extent of LR increases. This behaviour can be attributed to the impact of the outer LYSO crystal array reflective wrapping scattering the scintillation photons back into LYSO crystals residing within the 3$\times$3 SiPM pixel footprint. Overall, based on these observed trends, maximum light restriction to a 3$\times$3 SiPM pixel footprint can be achieve through increasing the PSI dilation and using LYSO crystal with a high surface roughness. 

\begin{figure}[tbh]    
    \centering
    \begin{subfigure}
    \centering 
        \includegraphics[width=0.235\textwidth, trim = {5 0 30 10},clip]{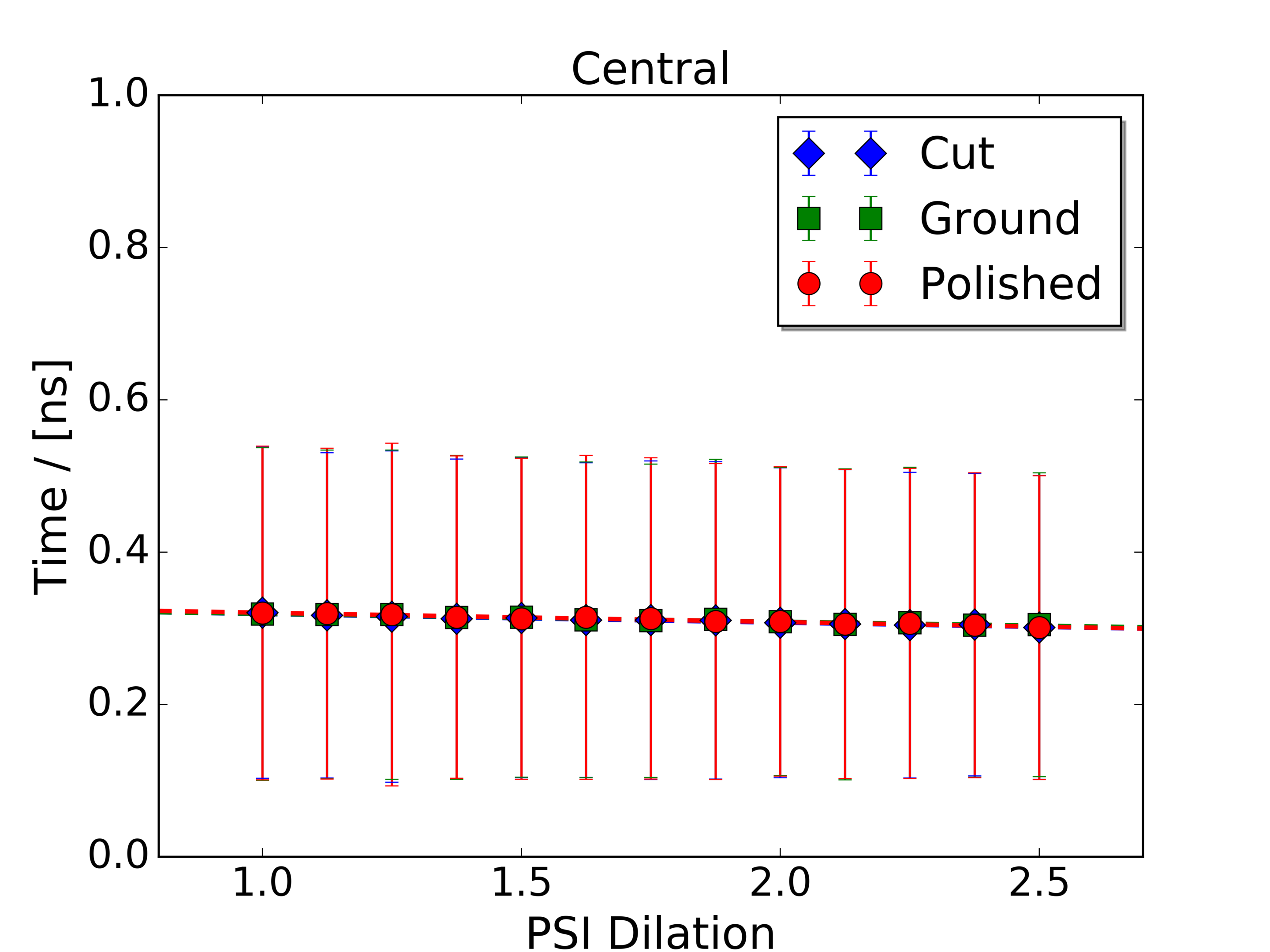}
        \label{fig:r5a}
    \end{subfigure}
    \begin{subfigure}
    \centering
        \includegraphics[width=0.235\textwidth, trim = {5 0 30 10},clip]{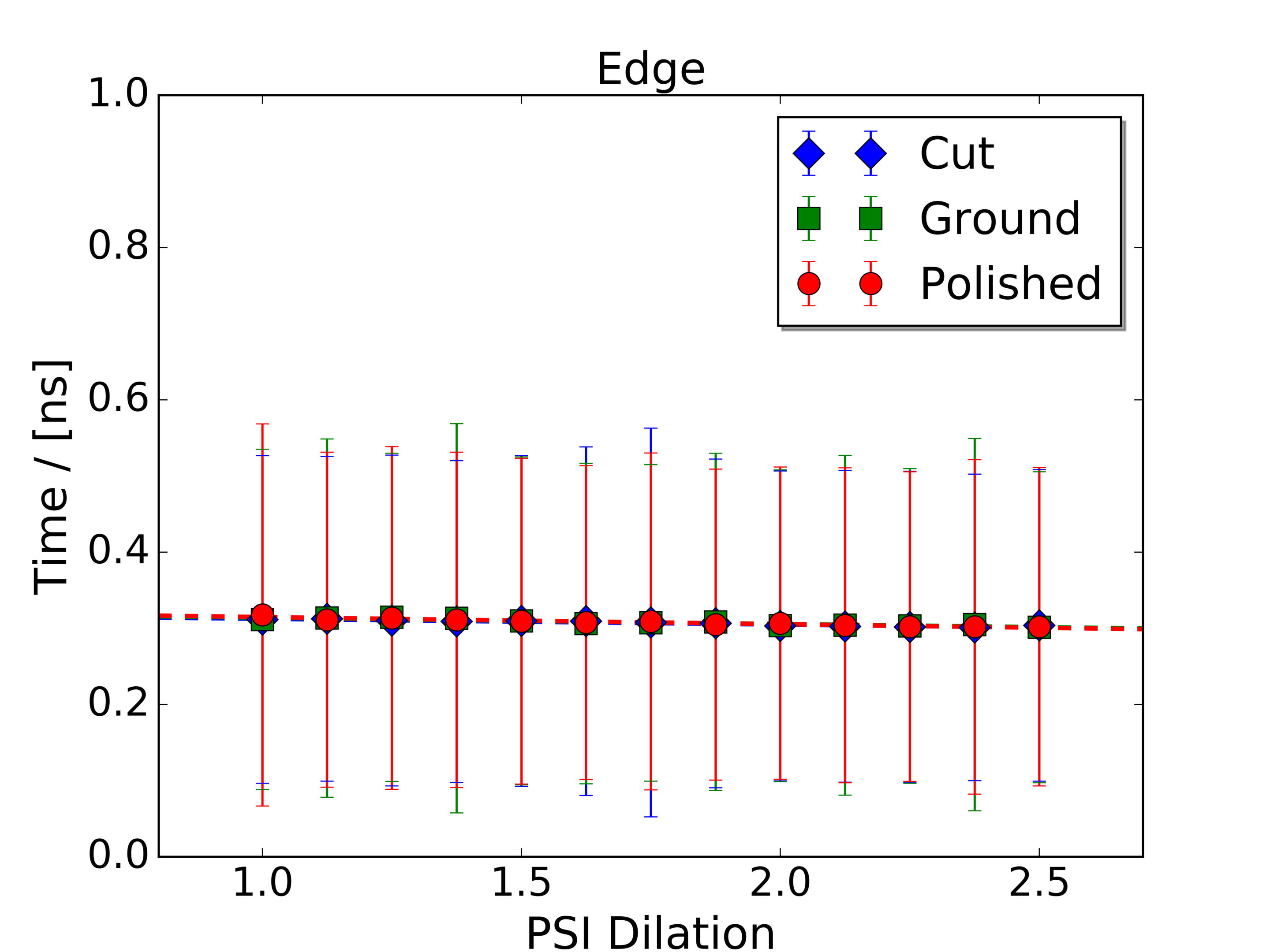}
        \label{fig:r5b}
    \end{subfigure}
    \newline
    \begin{subfigure}
    \centering 
        \includegraphics[width=0.235\textwidth, trim = {5 0 30 10},clip]{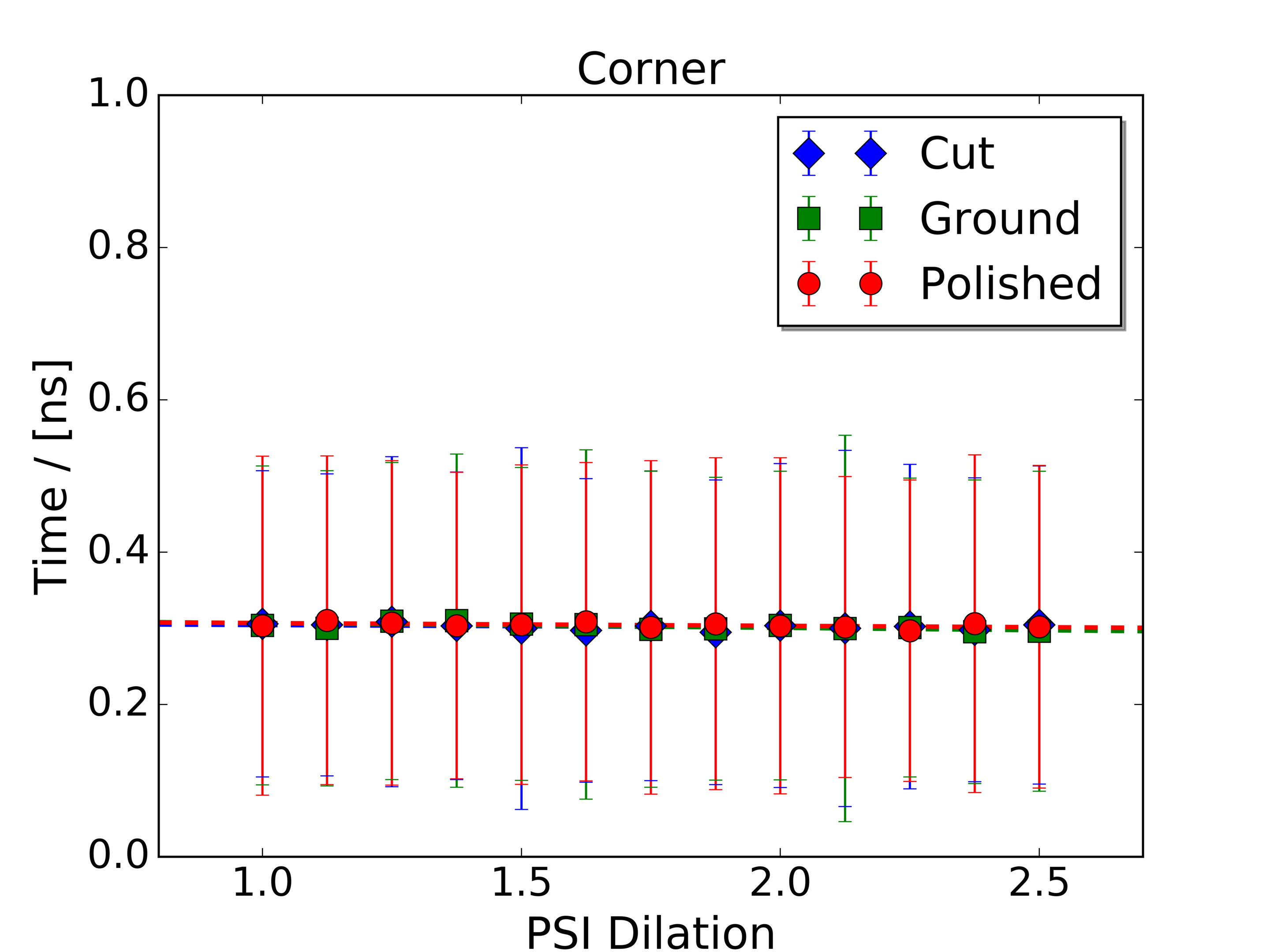}
        \label{fig:r5c}
    \end{subfigure}
    \begin{subfigure}
    \centering
        \includegraphics[width=0.235\textwidth, trim = {5 0 30 10},clip]{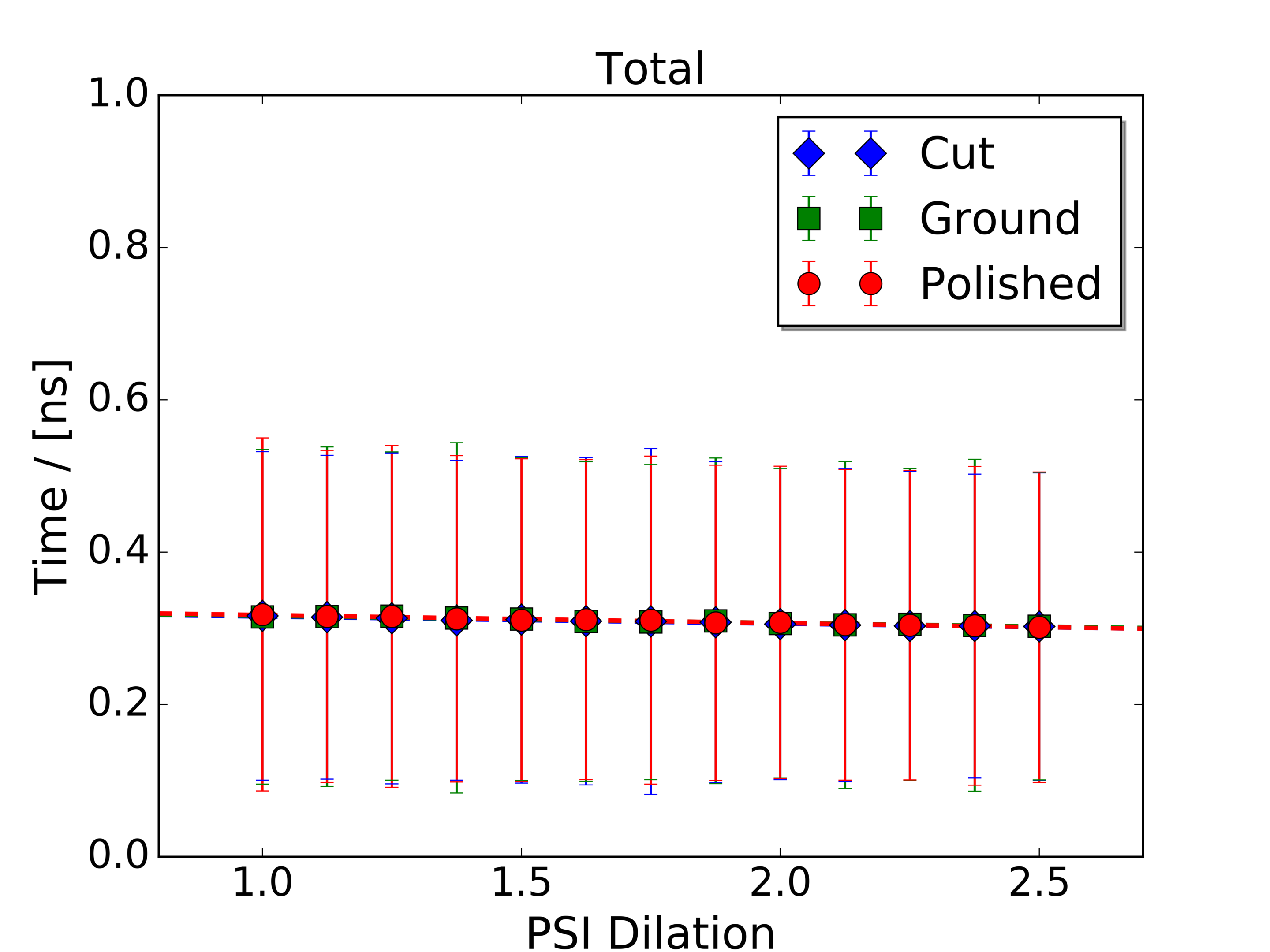}
        \label{fig:r5d}
    \end{subfigure}
\caption{Mean (markers) and standard deviations (bounding bars) of the 1st SPAD trigger relative to gamma-ray interaction time for four LYSO array crystal region classifications: central (top left), edge (top right), corner (bottom left), and total (bottom right). The coloured dash lines correspond to a fitted linear function for each crystal surface type to illustrate the general trend as a function of PSI.}
\label{fig:r5}
\end{figure}

\begin{figure}[tbh]    
    \centering
    \begin{subfigure}
    \centering 
        \includegraphics[width=0.235\textwidth, trim = {5 0 30 10},clip]{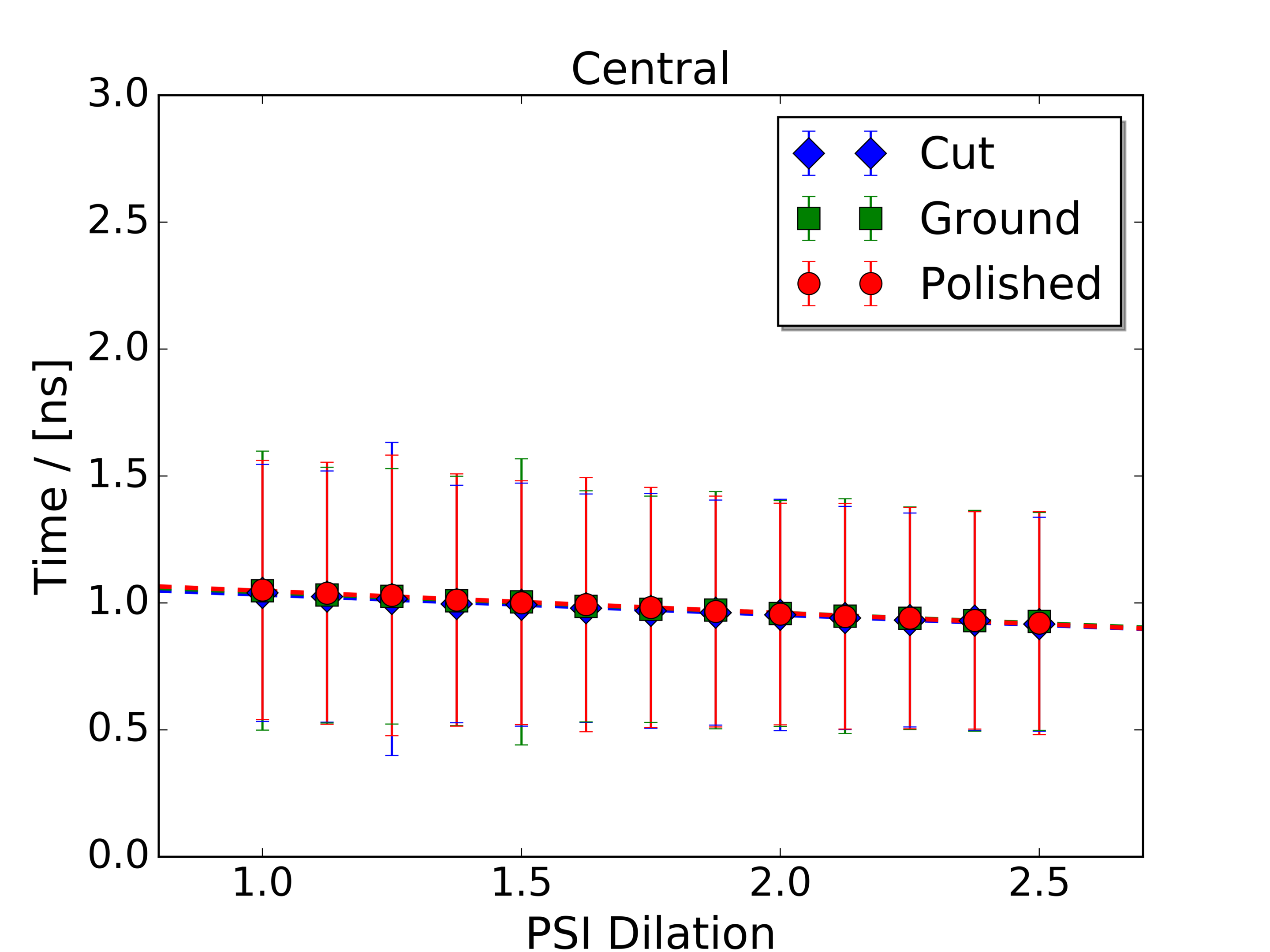}
        \label{fig:r6a}
    \end{subfigure}
    \begin{subfigure}
    \centering
        \includegraphics[width=0.235\textwidth, trim = {5 0 30 10},clip]{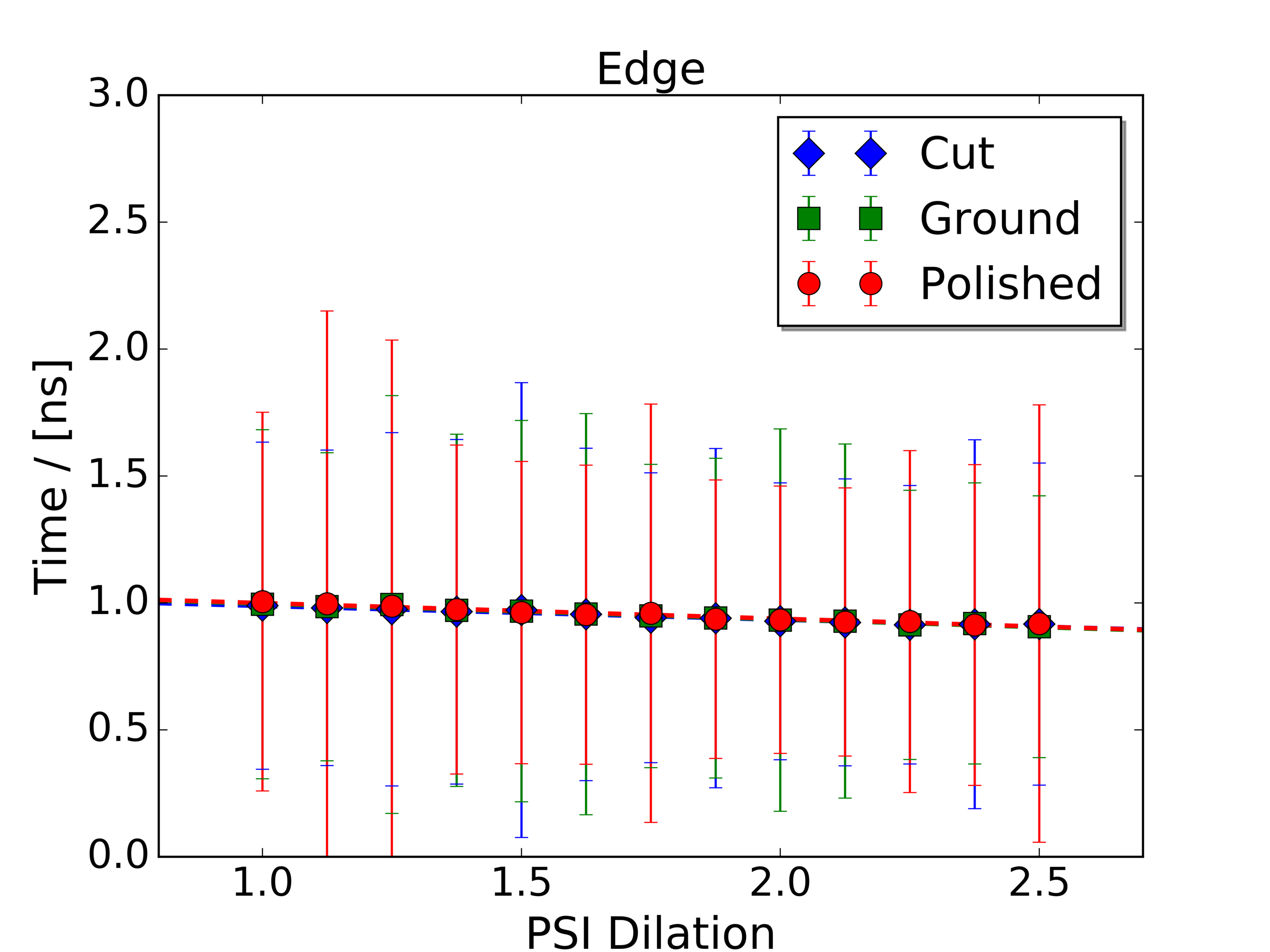}
        \label{fig:r6b}
    \end{subfigure}
    \newline
    \begin{subfigure}
    \centering 
        \includegraphics[width=0.235\textwidth, trim = {5 0 30 10},clip]{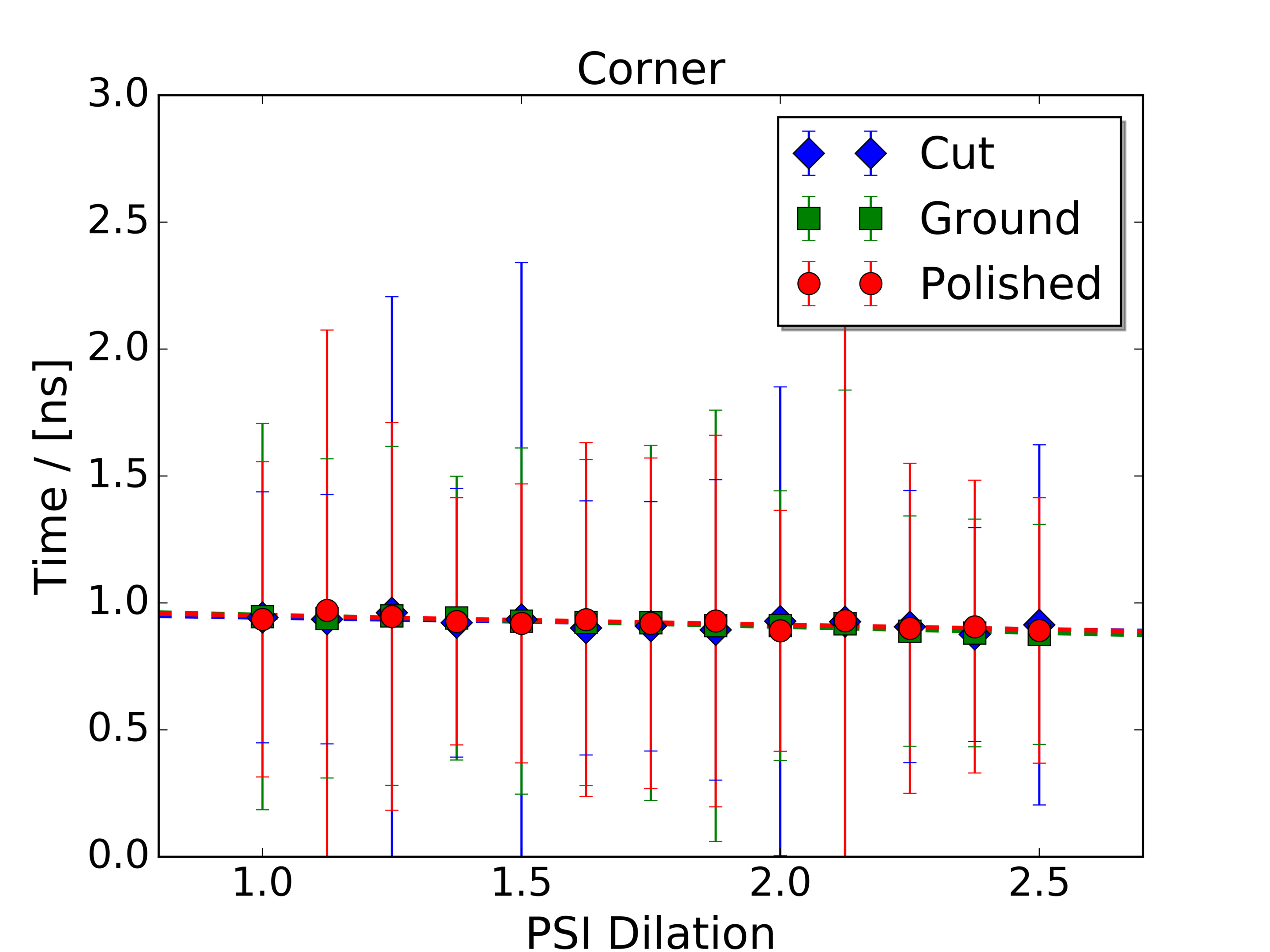}
        \label{fig:r6c}
    \end{subfigure}
    \begin{subfigure}
    \centering
        \includegraphics[width=0.235\textwidth, trim = {5 0 30 10},clip]{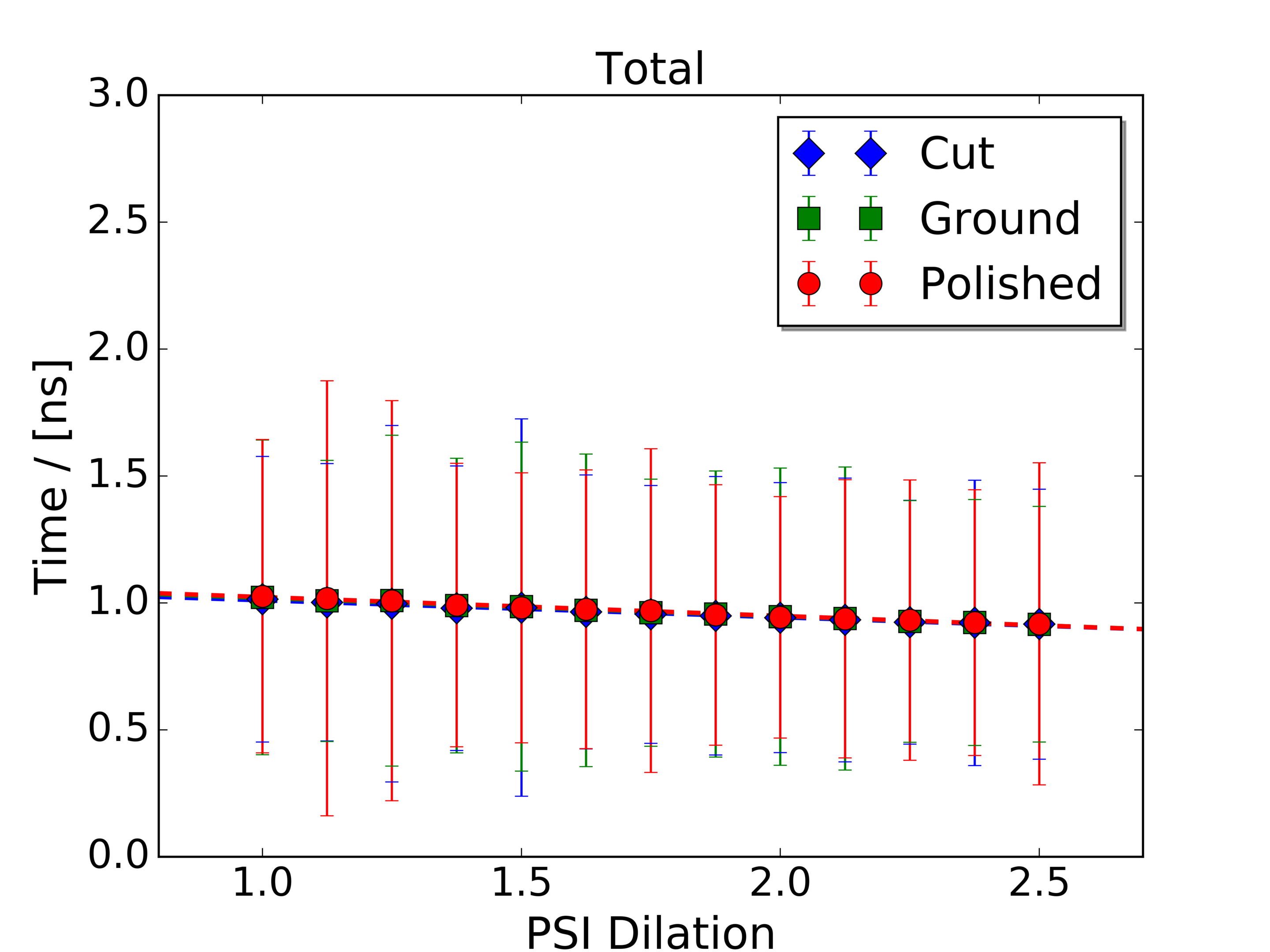}
        \label{fig:r6d}
    \end{subfigure}
\caption{Mean (markers) and standard deviations (bounding bars) of the 10th SPAD trigger relative to gamma-ray interaction time for four LYSO array crystal region classifications: central (top left), edge (top right), corner (bottom left), and total (bottom right). The coloured dash lines correspond to a fitted linear function for each crystal surface type to illustrate the general trend as a function of PSI.}
\label{fig:r6}
\end{figure}

\begin{figure}[tbh]    
    \centering
    \begin{subfigure}
    \centering 
        \includegraphics[width=0.235\textwidth, trim = {5 0 30 10},clip]{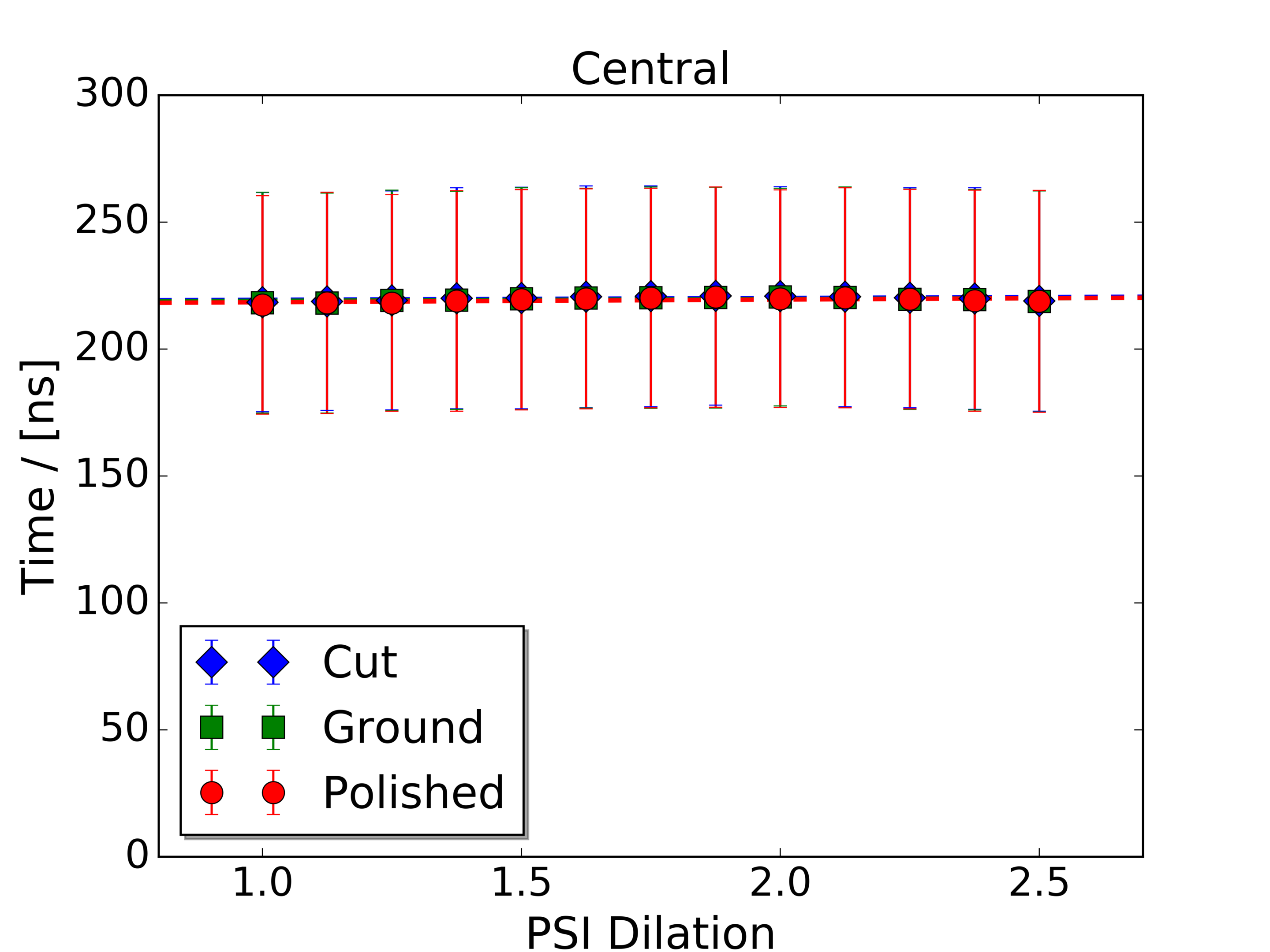}
        \label{fig:r8a}
    \end{subfigure}
    \begin{subfigure}
    \centering
        \includegraphics[width=0.235\textwidth, trim = {5 0 30 10},clip]{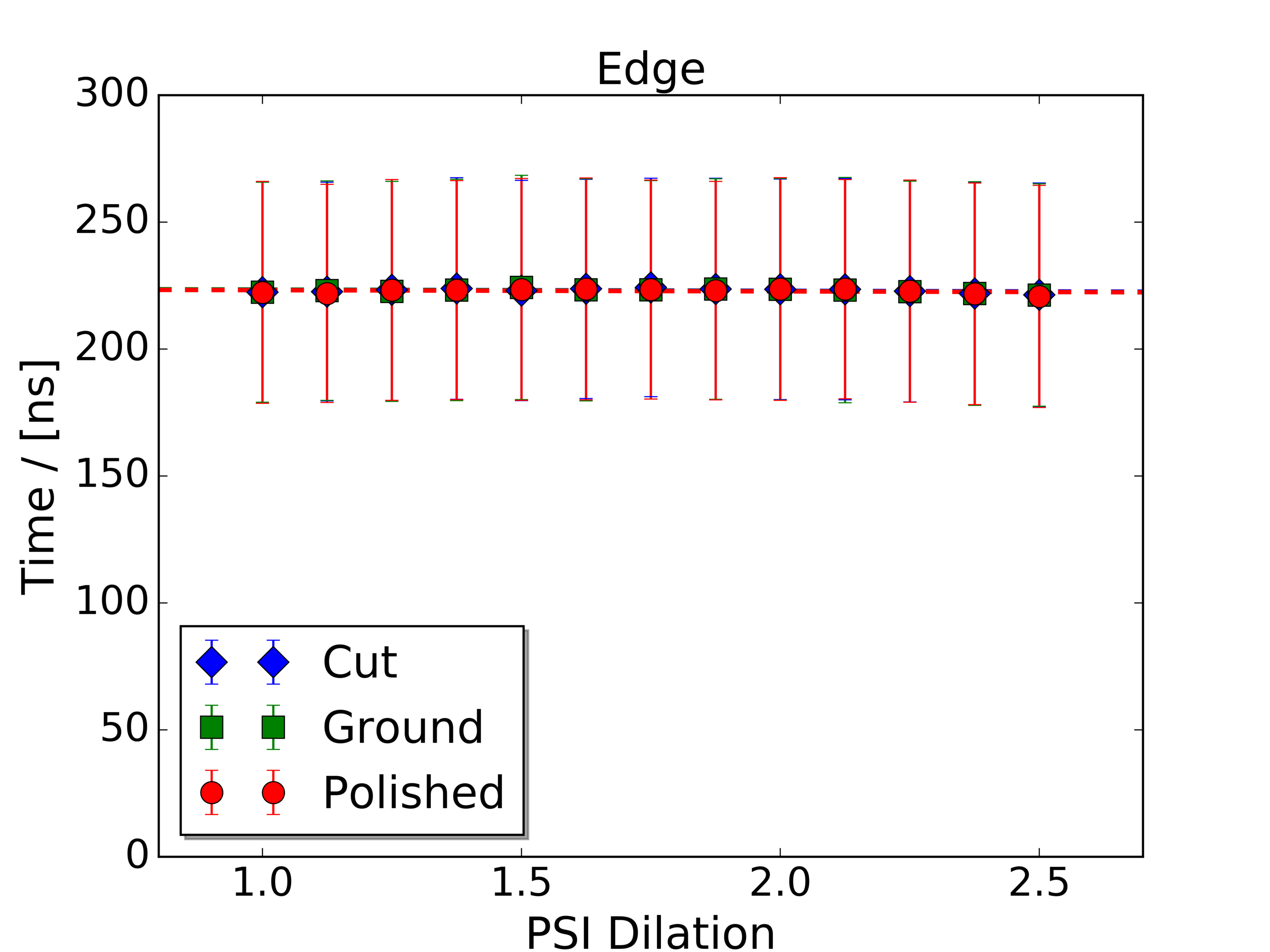}
        \label{fig:r8b}
    \end{subfigure}
    \newline
    \begin{subfigure}
    \centering 
        \includegraphics[width=0.235\textwidth, trim = {5 0 30 10},clip]{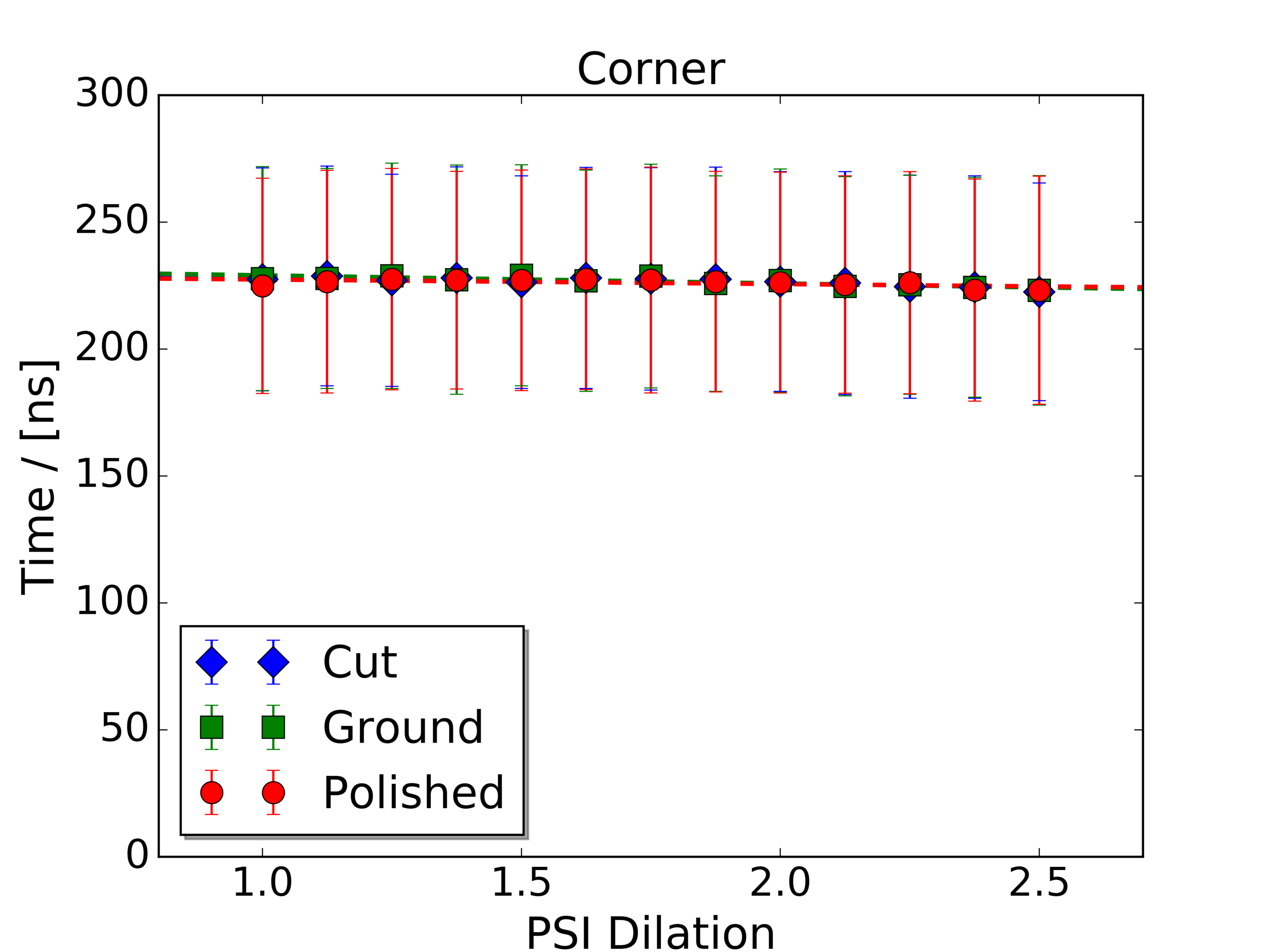}
        \label{fig:r8c}
    \end{subfigure}
    \begin{subfigure}
    \centering
        \includegraphics[width=0.235\textwidth, trim = {5 0 30 10},clip]{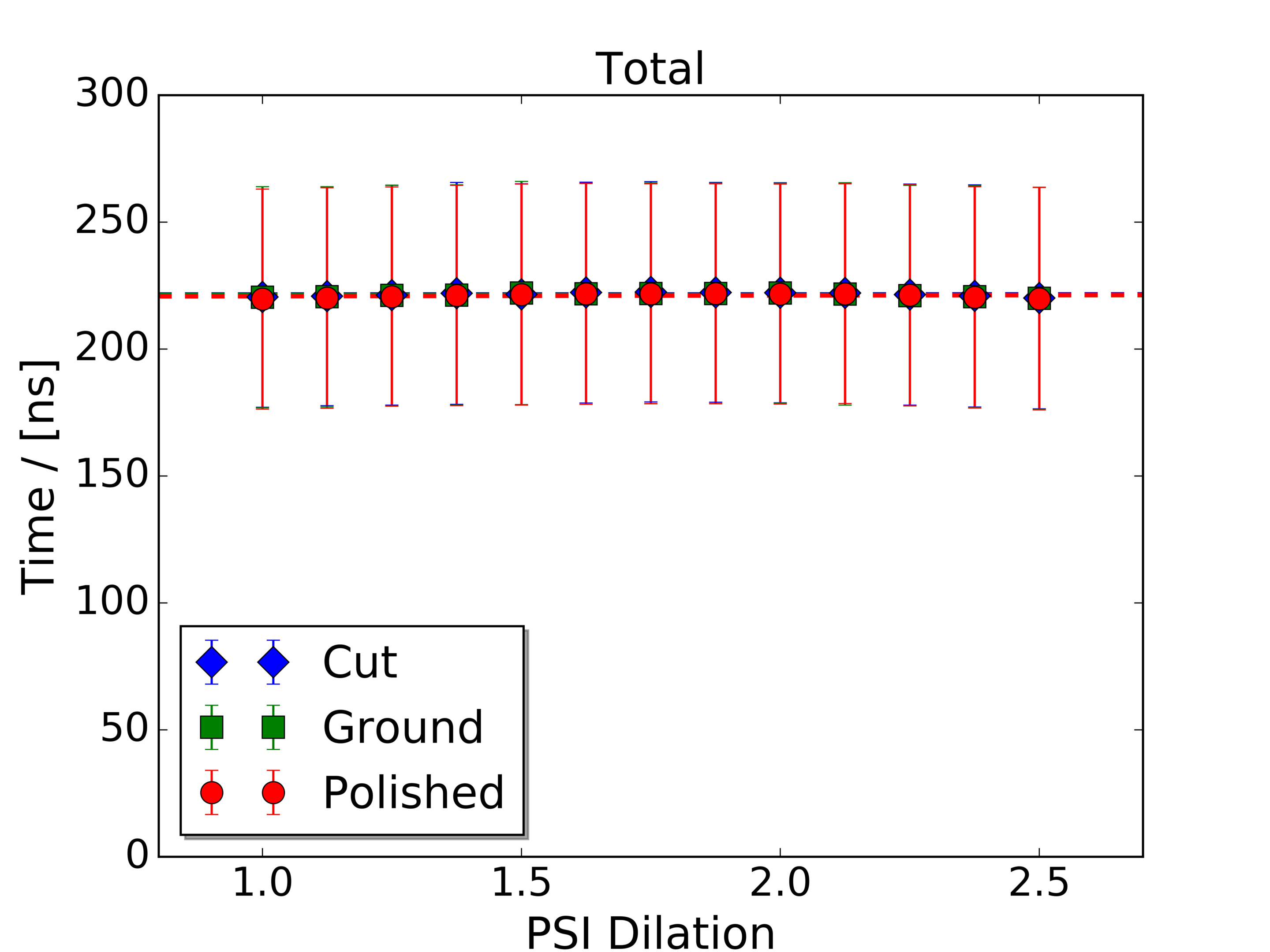}
        \label{fig:r8d}
    \end{subfigure}
\caption{Mean (markers) and standard deviations (bounding bars) of the final SPAD trigger relative to gamma-ray interaction time for four LYSO array crystal region classifications: central (top left), edge (top right), corner (bottom left), and total (bottom right). The coloured dash lines correspond to a fitted linear function for each crystal surface type to illustrate the general trend as a function of PSI.}
\label{fig:r8}
\end{figure}

The mean and standard deviation of the 1st, 10th and final SPAD trigger times for the different crystal surface types and LYSO crystal array region classifications can be seen in Figures \ref{fig:r5}, \ref{fig:r6}, and \ref{fig:r8} respectively. In these figures it can be seen that both the crystal surface roughness and crystal array region of gamma-ray interaction have minimal impact on mean time of the 1st, 10th and final SPAD trigger. For the impact of PSI dilation, there is a weak inverse relationship with respect to mean time of the 1st, 10th and final SPAD trigger for all explored crystal surface types and LYSO crystal array region classifications. These observed relationships are also true for the standard deviation of the 1st and final SPAD trigger times. However in the case of the standard deviation of the 10th SPAD trigger times, the trends for crystal surface roughness and PSI dilation hold true for the central, but not the edge and corner crystal array regions which could be attributed to the impact of scintillation photon scattering.

\section{Discussion}
\label{sec:D}
Assessment/optimisation of the performance of the proposed PET radiation detector design in the configurations outlined in Section \ref{sec:M} was undertaken through the use of five FoMs. Of these five FoMs, it was shown that for three of them (energy resolution, CoIIA, and mean/standard deviation in SPAD trigger time) that crystal surface roughness and foil PSI dilation had effectively zero impact. The two remaining FoMs, DoI and LR, displayed dependence on both the crystal surface roughness and foil PSI dilation. However, in the case of the impact of gamma-ray interaction location within the three different defined detector crystal array regions (central, edge, and corner), all but one of the FoMs followed the general trend that the central region possessed the best performance. This exception was for LR, where the edge and corner regions out performed the central region due to the impact of the outer LYSO crystal array reflective wrapping scattering the scintillation photons back into the 3$\times$3 SiPM pixel footprint region.

Of the two FoMs that observed a dependence on crystal surface roughness and foil PSI dilation, DoI and LR, their relative relationships are inverse to one another. Since these two FoMs can be linked to effective spatial resolution and maximum count rate before event pile-up, i.e. greated LR restriction would reduce the cross-talk between 3$\times$3 SiPM pixel footprint regions, it means that a native trade-off exists between these two crucial performance characteristics of the proposed PET radiation detector design \cite{Ito2011,Cherry2003,Bushberg2011}. For example to achieve the highest possible count rate before event pile-up, high crystal surface roughness and large foil PSI dilation would be required. Whereas to maximise the effective spatial resolution through increasing DoI accuracy, the opposite configuration would be required (e.g. polished crystals and minimal PSI dilation). Based on the data presented in Figures \ref{fig:r3} and \ref{fig:r4}, a compromise between the two could be achieved through the use of ground LYSO crystal and a PSI dilation of 1.75.

Within this work the photosensor electronic behaviour was treated in a simple manner, making it difficult to state any strong conclusion on the possible Time of Flight (ToF) performance \cite{Seifert2012,Tabacchini2014,Lecoq2017}. However, a rough estimate of the possible ToF performance without correction for DoI dependence can be drawn from the standard deviations seen in Figures \ref{fig:r5} and \ref{fig:r6} assuming that each DPC3200 was configure in such a manner to trigger on the 1st and 10th SPAD trigger respectively (here it is assumed that the uncertainties of two DPC3200 sum in quadrature, and the impact of Cherenkov emission is ignored \cite{Korpar2011,Brunner2013,Brunner2014}). Across the range of crystal surface roughness and foil PSI dilation explored, the mean standard deviations for the total LYSO crystal array was 210 and 470 ps for the 1st and 10th SPAD trigger time respectively. This would result in ToF FWHM times of 700 and 1600 ps assuming that the temporal profile resembled a Gaussian distribution. Whilst this performance would be acceptable for gamma-ray pair correlation, it is not sufficient enough to yield any substantial improvement in image quality from the implementation ToF line of response modulation in systems such as HYPMED \cite{Lewellen2008,Ito2011,Cherry2003,Bushberg2011}.

Finally, the CoIIA and DoI FoMs data illustrates that the implemented least squares readout approach would yield an approximate three dimensional spatial resolution of 2 to 2.5 mm. his result matches those obtained in Ito et al.'s in-silico investigation with their two-axis light sharing patterned reflector foil crystal array design and 1$\times$1$\times$16 mm$^{3}$ LSO crystals \cite{Ito2010}. Whilst a three dimensional spatial resolution of 2 to 2.5 mm would be acceptable for a standard clinical PET system \cite{Cherry2003,Bushberg2011}, it is insufficient for organ specific limited FoV PET inserts such as HYPMED which aim to achieve imaging resolutions on the order of 1 mm \cite{Lewellen2008,Ito2011}. Therefore, PET radiation detector design specific readout algorithms are need to maximise potential performance (e.g. advanced positioning algorithms \cite{Lee2015,VanDam2013,Schug2015,Muller2018,Muller2019}, DoI corrected ToF \cite{Lewellen2008,Ito2011}, etc.). This is a major consideration in the next phase of this work being undertaken at TUDelft, in which, an experimental prototype is being constructed utilising ground LYSO crystals and UV laser cut Vikuiti ESR foils of PSI dilation of 1.0 (produced by Micron Laser Technology\footnote{https://micronlaser.com/}) with the Philips DPC3200 photosensor. 

\section{Conclusion}
\label{sec:C}

To meet the PET detector performance requirements of organ-specific limited FoV PET/MR inserts, a novel two-axis patterned reflector foil pixelated scintillator crystal array design was developed and its proof-of-concept illustrated in-silico with the Monte Carlo radiation transport modelling toolkit Geant4. It was shown that the crystal surface roughness and phased open reflector cross-section patterns could be optimised to maximise either the PET radiation detector's effective spatial resolution, or count rate before event pile-up. In addition it was illustrated that these two parameters had minimal impact on the energy and time resolution of the proposed PET radiation detector design. Finally, it was determined that a PET radiation detector with balance performance could be constructed using ground crystals and phased open reflector cross-section pattern corresponding to the middle of the tested range. 

\appendices

\section*{Acknowledgements}
This project has received funding from the European Union’s Horizon 2020 research and innovation programme under grant agreement No 667211. It was also supported by the Multi-modal Australian ScienceS Imaging and Visualization Environment (MASSIVE) (http://www.massive.org.au). S. E. Brunner was supported by European Union’s Horizon 2020 Framework Programme, Marie Sklodowska-Curie action 659317 (PALADIN). 

\section{Geant4 In-Silico Test Platform Material Properties}
\label{appendix1}

\begin{figure}[tbh]    
    \centering
     \includegraphics[width=0.5\textwidth]{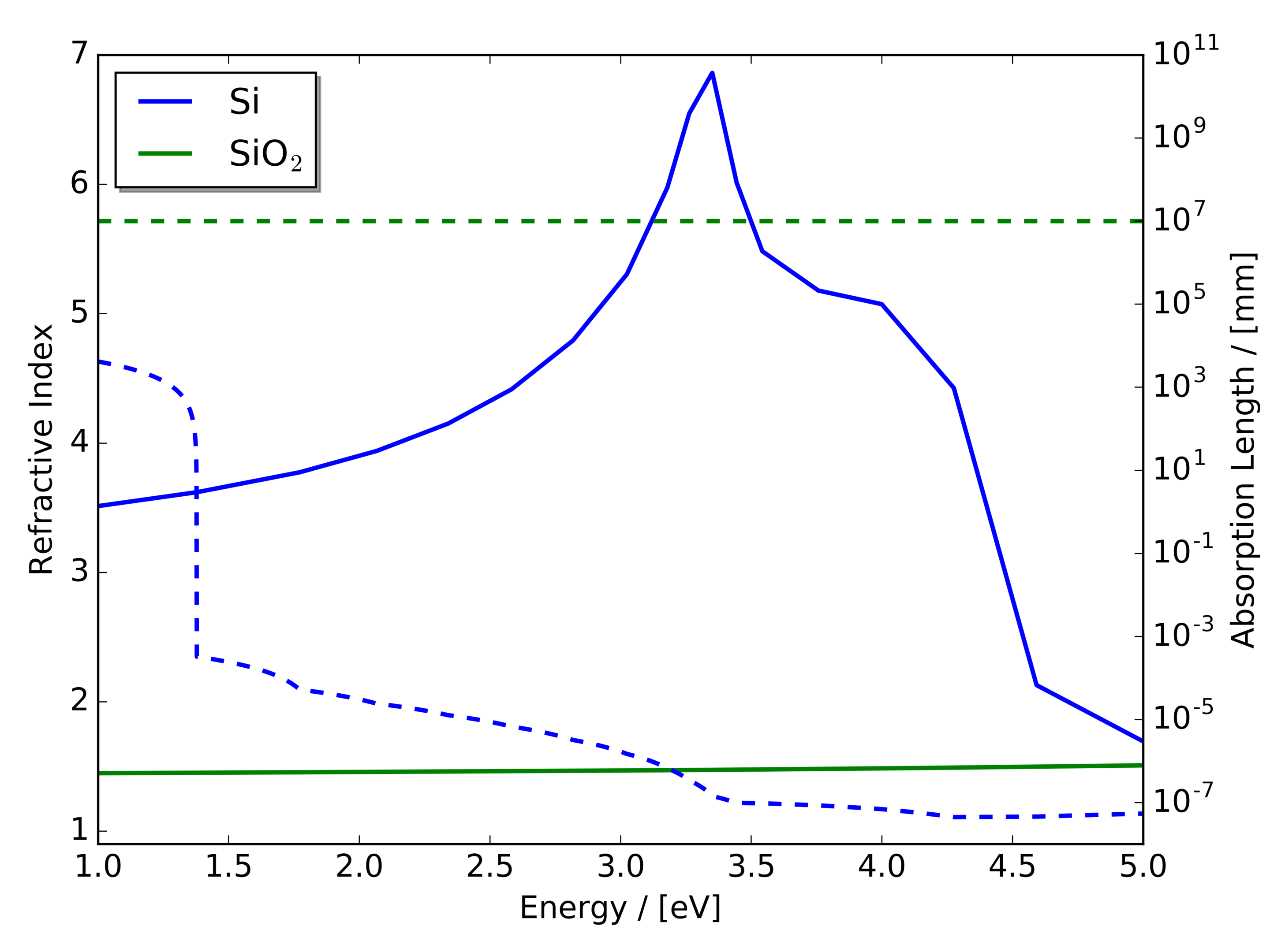}
\caption{DPC3200 pixel (Si) and quartz glass (SiO$_2$) material refractive index (solid line) and attenuation length (dashed line) data sets implemented in the Geant4 in-silico test platform.}
\label{fig:a1}
\end{figure}

\begin{figure}[tbh]    
    \centering
     \includegraphics[width=0.5\textwidth]{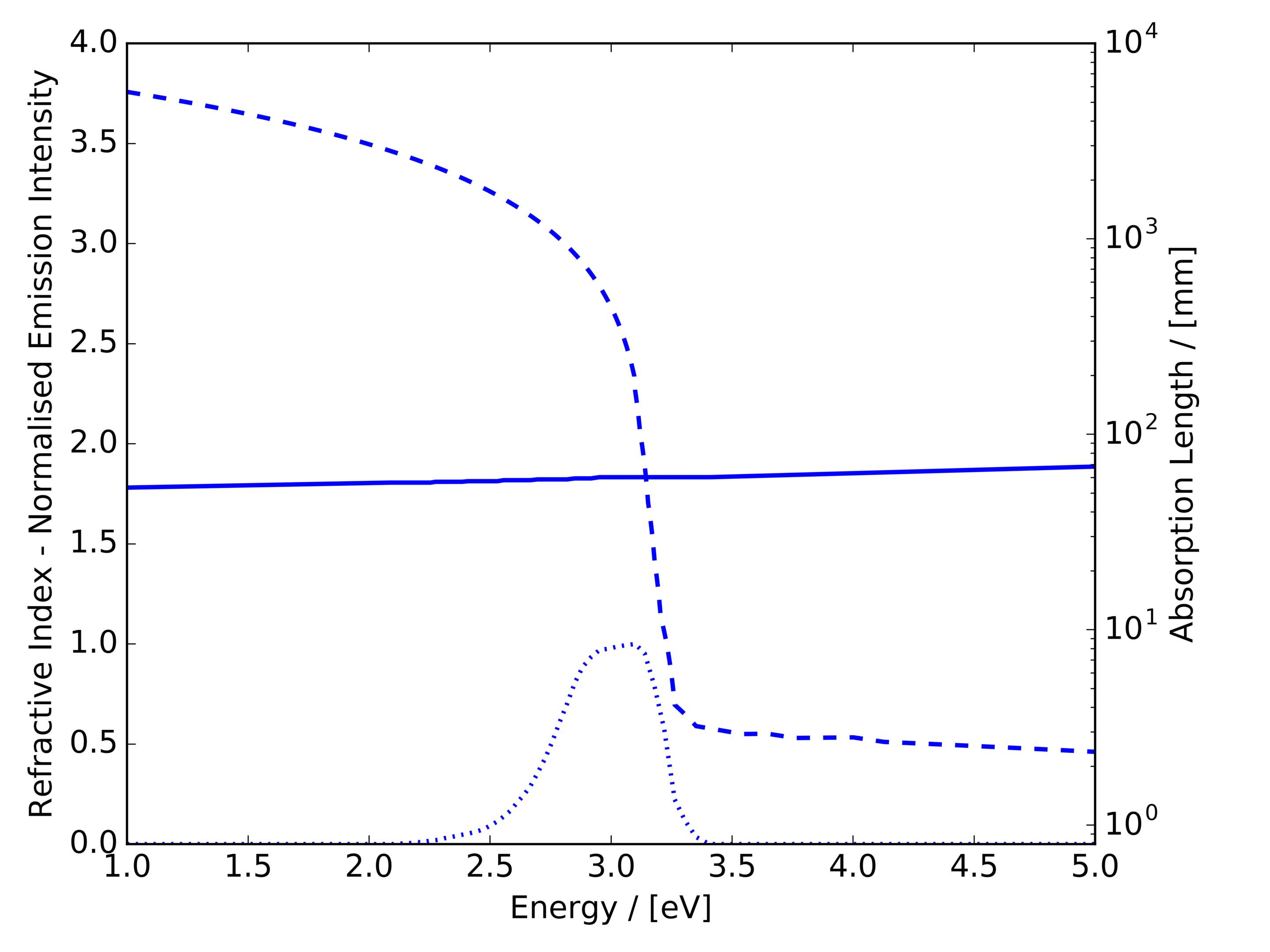}
\caption{LYSO scintillator crystal material refractive index (solid line), attenuation length (dashed line) and normalised scintillation photon emission intensity (dotted line) data sets implemented in the Geant4 in-silico test platform.}
\label{fig:a2}
\end{figure}

The following appendix contains the density, elemental composition, and optical/scintillation properties of all materials utilised in the developed Geant4 in-silico test platform. Material data relating to the world volume, Vikuiti ESR foil, bonding glue and implemented Philips DPC3200 SiPM is outlined in Table \ref{tab:1} and Figure \ref{fig:a1}. Whereas material data relating to the LYSO scintillator crystals, based on information from the Masters' thesis of Dachs \cite{Dachs2016}, can be seen in Table \ref{tab:2} and Figure \ref{fig:a2}.

\begin{table*}[h]
\centering
\begin{tabular}{||c|c|c|c|c|c||}
\hline
\hline
             &              &           &            & Optical        &          \\
 Material    & Density      & Elemental & Refractive & Reflectivity / & Reference \\
             & (g/cm$^{3}$) & Composition & Index      & Absorption   & \\
 \hline
 \hline             
 Air         & 1.29$\times$10$^-3$ & C (0.01\%), N (75.52\%), & 1 & - & Geant4 Material  \\
             &       &  O (23.19\%), Ar (1.28\%) &  &  &   Database \cite{G42016} \\
 DELO glue   & 1.0   & H$_8$C$_5$O$_2$ & 1.5   & -   & \cite{Dachs2016} \\
 Vikuiti ESR & 1.29  & H$_8$C$_{10}$O$_4$& -     & 98\% / 2\%  & \cite{3M2019} \\
 DPC3200 PCB & 1.86  & SiO$_2$ (52.8\%), H$_1$C$_1$O$_1$ (47.2\%) & - & 0\% / 100\% & \cite{Dachs2016} \\
 DPC3200 Pixel &  2.33 & Si & See Figure \ref{fig:a1} & See Figure \ref{fig:a1} & \cite{Philipp1960} \\
 DPC3200 Glass & 2.203 & SiO$_2$ & See Figure \ref{fig:a1} & See Figure \ref{fig:a1} & \cite{Dachs2016} \\
 \hline   
 \hline
\end{tabular}
\\
\caption[]{Density, elemental composition, and optical material properties of the world volume, Vikuiti ESR foil, bonding glue and Philips DPC3200 SiPM implemented in the Geant4 in-silico test platform.}
\label{tab:1}
\end{table*}

\begin{table*}[h]
\centering
\begin{tabular}{||c|c|c|c|c|c|c||}
\hline
\hline
             &           &             & Optical Yield,           &   Optical Decay   &                   &           \\
 Density      & Elemental & Refractive & Emission Spectrum,        &  Time Constants  & Resolution Scale  & Reference \\
(g/cm$^{3}$) & Composition & Index      & Absorption Length         &       (ns)      &  (at 511 keV)     & \\

 \hline
 \hline      
  7.4       & Lu$_{1.9}$Y$_{0.1}$Si$_1$O$_5$ &  See Figure \ref{fig:a2} & 30 Photons per eV, & Fast: 7.1 (7\%) & 4.17  & \cite{Dachs2016} \\
            & (0.5\% Ce doping)           &                          & See Figure \ref{fig:a2} & Slow: 33.3 (93\%) &   & \\
 \hline   
 \hline
\end{tabular} 
 \caption[]{Density, elemental composition, and optical properties of the LYSO material implemented in the Geant4 in-silico test platform.}
\label{tab:2}
\end{table*}

\end{document}